\documentclass[journal,10pt]{IEEEtran}
\usepackage{color}
\usepackage{cite}

\ifCLASSINFOpdf
   \usepackage[pdftex]{graphicx}
\else
   \usepackage[dvips]{graphicx}
\fi
\usepackage[cmex10]{amsmath}
\usepackage[noend]{algpseudocode}
\usepackage{algorithm}
\makeatletter
\def\BState{\State\hskip-\ALG@thistlm}
\makeatother

\usepackage{booktabs}

\usepackage{array}
\usepackage{url}

\begin{document}
%
\title{Improved Utility-based Congestion Control for Delay-Constrained Communication}
%
%
%

\author{Stefano~D'Aronco,~
	    Laura~Toni,~
	    Sergio~Mena,~
	    Xiaoqing~Zhu,~
        and~Pascal~Frossard~}
\maketitle

\begin{abstract}
Due to the presence of buffers in the inner network nodes, each congestion event leads to buffer queueing and thus to an increasing end-to-end delay. In the case of delay sensitive applications, a large delay might not be acceptable and a solution to properly manage congestion events while maintaining a low end-to-end delay is required.  Delay-based congestion algorithms are a viable solution as they target to limit the experienced end-to-end delay. Unfortunately, they do not perform well when sharing the bandwidth with congestion control algorithms  not regulated by delay constraints (e.g., loss-based algorithms). Our target is to  fill this gap,  proposing a novel congestion control algorithm for delay-constrained communication over best effort packet switched networks. The proposed algorithm is able to maintain a bounded queueing delay when competing with other delay-based flows, and avoid starvation when competing with loss-based flows.
We adopt the well known price-based distributed mechanism as congestion control, but \emph{i)} we introduce a novel  non-linear mapping between the experienced delay and the price function, and \emph{ii)} we combine both delay and loss information into a single price term based on packet interarrival measurements. We then provide a stability analysis for our novel algorithm and we  show its performance in the simulation results carried out in the NS3 framework. Simulation results demonstrate that the proposed algorithm is able to: achieve good intra-protocol fairness properties, control efficiently the end-to-end delay, and finally, protect the flow from starvation when other flows cause the queuing delay to grow excessively.
\end{abstract}


%
\IEEEpeerreviewmaketitle

\section{Introduction}
%
%
%
%

\IEEEPARstart{N}{owadays,} many Internet applications aim to work not only at maximizing their throughput, but also at meeting crucial delay constraints in the transmission of data flows. Video conferencing applications are a good example of such delay sensitive services, where an excessive playback delay with the audio/video stream can drastically affect the quality of an internet call. On-line video gaming and desktop remote control are other examples of applications that require low latency communications.
When network links are congested these applications need to adjust their sending rate such that the experienced one-way delay is kept low and bounded, while preserving fairness with other flows. This reduces to a constrained resource allocation problem, which has to be solved in a fully distributed manner due to scalability issues.

Congestion control algorithms can be seen as distributed methods to solve optimal network resource allocation problems. These algorithms are usually categorized in primal or dual algorithms, based on the solving method adopted~\cite{kelly,surveyint,paganini}.
From a more practical point of view, the primal and dual congestion control algorithms roughly correspond, though not exactly, to loss-based and delay-based algorithms. Loss-based controllers are widely deployed over the internet (e.g., TCP) and use congestion events triggered by packet losses to perform rate adaptation. However this class of controllers does not take into account any type of delay measurement, such as the One-Way Delay (OWD) or the Round Trip Time (RTT). Hence, there is no control on the latency that the packets might experience on their route and large delays can be experienced in the case of long buffers in the inner network nodes.  
On the other hand, delay-based congestion control algorithms can overcome the increasing delay issue by detecting congestion events from OWD measurements. Delay-based congestion control algorithms are therefore suitable for low delay applications since they are able to keep a low communication delay by adapting the sending rate to the evolution of the delay.
However, they usually suffer from poor performance when sharing the network with loss-based controllers. 
Loss-based congestion control  algorithms always fill the buffers of the inner network nodes before triggering congestion events. Therefore, any delay-based flow sharing the same bottleneck might experience a too large queuing delay and quickly reach starvation (i.e., a sending rate close to zero).
There is the need for a congestion control that could enable low delay communication whenever possible (essential for low delay communication) and that is robust against the presence of loss-based flows (mostly deployed in the internet)

Beyond this coexistence challenge, new congestion control algorithms to be deployed in the Internet need: $i)$ to provide good inter-protocol performance when competing against existing controllers, such as TCP; $ii)$ to mainly act at the endpoints rather than at the inner network nodes (modifications of the inner network nodes are particularly difficult); $iii)$ to be robust to noisy measurement of network parameters (i.e., propagation delay).
Many congestion control algorithms have actually been proposed in the past, but to the best of our knowledge there is no delay based congestion control algorithm able to well perform in the presence of loss-based flows and satisfying at the same time the three main implementation challenges listed above.

\begin{figure}
\centering
\includegraphics[scale=0.24]{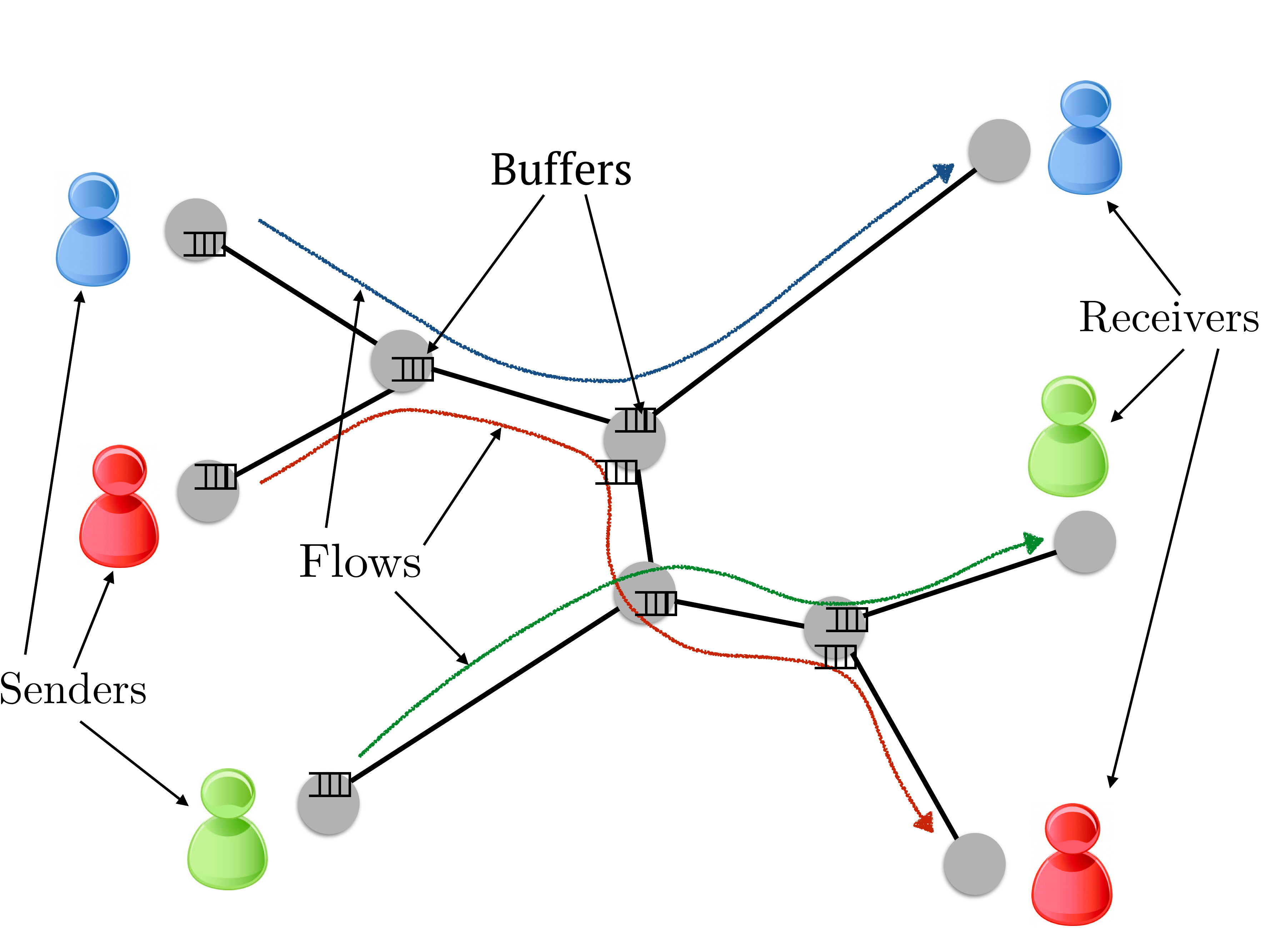}
\caption{Network system model.}
\label{modelfig}
\end{figure}

In this work, we target to fill this gap by proposing a new distributed Delay-Constrained Congestion Control (DCCC) algorithm  that is able to adapt the sending rate to both loss and delay-based congestion events and to overcome the aforementioned  issues. The ultimate objective is to preserve the low end-to-end delay constraint that is imposed by the application, when competing with other delay-based controlled flows, and at the same time, avoid starvation when competing with loss-based flows.
In more details, we consider a scenario where users send delay-sensitive data over a packet switched network, as shown in Fig.~\ref{modelfig}. The network is composed of a set of links and nodes, with the links being shared among different users who set up unicast communications between two endpoints of the network.
The proposed controller measures the experienced OWD and the interarrival time of the received packets at the receiver node, and adjusts the rate accordingly in order to maximize the overall utility of the network flows. The key intuition is that the interarrival time of the packets is correlated to both losses and queueing delay variations. Hence, by using this metric, the controller is able to work in both delay-based and loss-based environments. The ability to avoid starvation when competing against loss-based flows, and still guarantee a bounded experienced delay is made possible by the use of a non-linear mapping between the experienced OWD and the penalty congestion signal used by the rate update equation.
The DCCC algorithm has been implemented in the NS3 network simulator and has been tested under different topologies and working conditions. Simulation results show the ability of the proposed algorithm to keep bounded the value of the experienced OWD, to achieve a good intra-protocol fairness and to avoid starvation when competing with loss-based flows, such as TCP.
Note that the proposed algorithm relies on the OWD measure, which is meaningful only in the case of synchronized endpoints. However, the controller can still operate properly in the case of unsynchronized endpoints if the desynchronization is relatively small with respect to the OWD experienced.

The remainder of this paper is organized as follows. In Section~\ref{NUM}, we provide a description of the system model and the Network Utility Maximization (NUM) framework that is used in this paper. In Section~\ref{ctrl}, we describe our new congestion control algorithm and we analyze its stability in Section~\ref{stability}. In Section~\ref{simulation}, we provide some last details about the correct implementation of the controller and present the simulation results. Section~\ref{related} describes the related work. Finally conclusions are provided in Section~\ref{conc}.

\section{Network Utility Maximization}\label{NUM}

\subsection{System Model}\label{model} 
We consider a system with a set of $R$ users (or flows) indexed by $r \in R$ sharing a set of network resources. Each user transmits data at a sending rate $x_r$, in a single-path unicast communication between two endpoints. Let $l \in L$ be the links of the network, and $\mathcal{R}$ be  the $L \times R$ routing matrix, where $\mathcal{R}_{lr}=1$ if link $l$ is used by user $r$ and zero otherwise. The total rate passing through link $l$, denoted by $y_l$, can then be defined as
\begin{equation}
	y_l=\sum_{r \in R} \mathcal{R}_{lr} x_r.
	\label{eq:ratelink}
\end{equation}
Every link has a fixed maximum channel capacity $c_l$, which is constant over time. Each link is preceded by a buffer that accumulates the data in surplus when $y_l$ exceeds the link capacity $c_l$. The queueing delay at the buffer of link $l$, $q_l(t)$,  evolves over time as follows:
\begin{equation}
	\dot{q}_l = \ \frac{1}{c_l} \left( y_l - c_l \right)_{q_l}^+,
	\label{eq:link_q}
\end{equation}
Where $(z)_x^+$ is a projection that keeps the queuing delay positive. Here, as in the remainder of the work, we identify by $\dot{z}$ the derivative of the variable $z$ with respect to time.  A data unit, or packet, that arrives at link $l$ at time $t$ takes a time equal to $q_l(t)$ before being actually sent over the link. We further denote by $p_l$ the propagation delay of link $l$, which is defined as the time required for the data to propagate through the physical link; we assume $p_l$ to be constant over time. Therefore the total time for a data unit to traverse the link $l$ is given by the sum of the propagation and queueing delays, i.e., $d_l=p_l+q_l$.
The total one-way delay experienced by user $r$, $e_r$, is given by the sum of all the delays encountered on the path that connects the source node and the receiver node, which can be mathematically defined as
\begin{equation}
	e_r=\sum_{l \in L} \mathcal{R}_{lr} d_l.
\end{equation}
Finally, the network buffers have a maximum size $q_l^{\text{MAX}}$ and a droptail policy is implemented. This means that when the queue length reaches $q_l^{\text{MAX}}$, the next incoming packets are discarded. The loss ratio for link $l$ when its buffer is full corresponds to:
\begin{equation}
	\pi_l = \ \left( \frac{y_l-c_l}{y_l} \right)_{y_l-c_l}^+,
	\label{eq:loss_ratio}
\end{equation}
and the total loss ratio for user $r$ is:
\begin{equation}
	\pi_r = 1 - \prod_{l \in L} \mathcal{R}_{lr} \left(1-\pi_l \right),
	\label{eq:loss_ratio2}
\end{equation}
which, if the loss ratios of the links are small, can be approximated by
\begin{equation}
	\pi_r \simeq \sum_{l \in L} \mathcal{R}_{lr} \pi_l.
	\label{eq:loss_ratio3}
\end{equation}
We further introduce another notion of communication delays between the network nodes. We denote by $\tau_{rl}^f$ the delay experienced by data packets from the source node of user $r$ to the inner node located at the end of the link $l$. Note that $\tau_{rl}^f$ is defined only if link $l$ is a link of the route of flow $r$, and $\tau_{rl}^f=e_r$ if $l=L_r$, with $L_r$ being the final link of route $r$. Introducing the timing aspects in Eq.~\eqref{eq:ratelink} leads to have that the total rate of link $l$ at time $t$ is a function of the sending rates at time $t-\tau_{rl}^f$, this means\footnote{Note that in case of losses Eq.~\eqref{eq:ratelink2} and Eq.~\eqref{eq:ratelink} do not hold exactly due to the actual rate decrease caused by packet drops. The equations can however still be considered valid in low-loss regime.}:
\begin{equation}
	y_l(t)=\sum_{r \in R} \mathcal{R}_{lr} x_r(t-\tau_{rl}^f).
	\label{eq:ratelink2}
\end{equation}
Similarly we define  $\tau_{rl}^b$ as the time needed for the data to travel from link $l$ to the sink node and then to be sent back to the source node $r$. Note that any kind of network event is first detected by the receiver node and then is sent back to the source node, therefore the following condition holds:
\begin{equation}
	\text{RTT}_{r}=\tau_{rl}^f+\tau_{rl}^b,\ \ \forall l \in r.
	\label{eq:RTT}
\end{equation}
The $\text{RTT}_r$ of user $r$ can also be expressed as a function of the experienced user delay:
\begin{equation}
	\text{RTT}_{r}=e_r+e^b_r,
	\label{eq:RTT_2}
\end{equation}
where $e^b_r$ is the total experienced delay of the backward path. In the cases where the backward path is not congested, it is reasonable to assume that $e^b_r$ is equal to the propagation delay of the forward path. In any case, we assume that the backward delay $e^b_r$ does not depend on the rate $x_r$.

We introduce now the concept of utility function. We denote by $U_r(x_r)$ the benefit for user $r$ of sending data at rate $x_r$. The utility function is usually considered to be a concave increasing function of the rate. In the following we consider a logarithmic utility function given by:
\begin{equation}
	U_r(x_r)=h \log(x_r),
	\label{eq:log_util}
\end{equation}
where $h$ is a simple weight parameter that can take different values depending on the user $r$, and thus modify the importance of the different flows. Finally, we point out, that our framework can be extended to any differentiable increasing concave utility function, leading possibly to different results.

\subsection{Optimization Problem}\label{problem} 
We now define the NUM problem based on the framework of~\cite{kelly}. 
The goal is to optimally allocate the rates of the users in such a way that the overall utility is maximized and that the capacity constraints of the network links are respected.
The NUM problem is typically defined as follows:
\begin{equation}
	\begin{aligned}
	\text{NUM: } & \underset{x}{\text{maximize }}  \sum_{r \in R} U_r(x_r) & \\
	& \text{subject to}   \sum_{l \in L} \mathcal{R}_{lr} x_r \leq c_l, \  l = 1, \ldots, L. &
	\end{aligned}
	\label{opt_1}
\end{equation}
The NUM problem can ideally be solved exactly in a centralized way. In this case the exact solution would be first computed and then communicated to the users. As a result the capacity constraints would never be violated, leading to no losses and no queueing delay at the bottlenecks links. The requirement of a prior full knowledge of the network state prevents this solving method to be actually usable. Thus, the NUM problem is typically decoupled and then solved in a distributed manner.

\subsection{Solution by Decomposition}\label{decomp} 
We discuss now the most commonly used approaches to solve the NUM problem, which include solutions based on penalty decomposition and on dual decomposition, and we point out their main limitations. We refer to~\cite{tutorial}  for a detailed tutorial on decomposition methods for NUM. A viable method is by using penalty functions, or prices, which map the violation level of the capacity constraints into a negative utility. The problem in~\eqref{opt_1} can be reformulated as follows:
\begin{equation}
	\underset{x}{\text{maximize }}  \sum_{r \in R} U_r(x_r) - \sum_{l \in L} \int_0^{y_l} g_l(z)dz
	\label{opt_pen}
\end{equation} 
where $g_l(z)$ is seen as the price to pay for using link $l$ when the link rate is equal to $z$. It is important for $g_l(\cdot)$  to be a positive and increasing function of the link rate. The problem in Eq.~\eqref{opt_pen} can be solved by a gradient-based algorithm with the following the rate update equation:
\begin{equation}
	\begin{aligned}
	\dot{x}_r & = \alpha_r \left( U'_r(x_r) - \sum_{l \in L} \mathcal{R}_{lr} g_l(y_l)\right),\\
	\end{aligned}
	\label{opt_pen_dyn}
\end{equation} 
where $U'_r(x_r)$ is the derivative of the utility function with respect to $x_r$. In the reminder of the paper the notation $f'(z)$ denotes the derivative of the function $f(z)$ with respect to a general variable $z$.
Imposing the loss ratio of link $l$ as price: $g_l(y_l)=\pi_l(y_l)$, and assuming $\mathcal{R}_{lr} \pi_l \simeq \pi_r$, the problem is decoupled since users need to know only their own utility function and their loss ratio $\pi_r$ to compute the rate update. Users can easily estimate their loss ratios by observing the number of dropped packets at the receiving node. The main drawback of this solution is that at the equilibrium losses are always experienced, which means that bottleneck buffers are always full and large queuing delays might be experienced.
A large family of congestion control algorithms, such as the classical loss-based version of TCP,  can be modeled as systems governed by Eq.~\eqref{opt_pen_dyn}.

Another possible way to solve the NUM problem is by dual decomposition.  The optimization problem can be solved in a distributed way by using a primal-dual algorithm based on the following update equations:
\begin{IEEEeqnarray}{rCl}
	\label{opt_dual_dyn}
	\dot{x}_r & = &\alpha_r \left( U'_r(x_r) - \sum_{l \in L} \mathcal{R}_{lr} \lambda_l \right) \IEEEyessubnumber \label{opt_dual_dyn_1} \\
	\dot{\lambda_l}  & =  &\nu_l \left( y_l - c_l \right)_{\lambda_l}^+, \IEEEyessubnumber \label{opt_dual_dyn_2}
\end{IEEEeqnarray}
where $\lambda_l$ is the dual variable associated to the capacity constraint of link $l$, and
$\nu_l$ is a simple positive parameter that controls the rate of update of the dual variable $\lambda_l$.
Note that if we set $\nu_l=1/c_l$, the dual variables have exactly the same form of the queueing delays defined in Eq.~\eqref{eq:link_q}.
Therefore, users can compute Eq.~\eqref{opt_dual_dyn_1} by estimating the total queueing delay of their own route, and adapt their rates accordingly.
For a given utility function the delay experienced at the equilibrium of a user depends on the bottleneck capacity. Generally, the larger the bottleneck capacities $c_l$, the larger the user rates $x_r$ at equilibrium, and the lower the delays at equilibrium.
It is worth noting that a flow that converges at the equilibrium to a rate equal to $x$, needs to experience (create) a certain amount of queueing delay along its route, which is usually called self-inflicted delay.

The basic dual decomposition method of Eq.~\eqref{opt_dual_dyn} has two main limitations: $i)$ it is not robust against loss-based flows, as already explained in the introduction; $ii)$ it assumes a perfect knowledge of the total queueing delay (when $\lambda_l=q_l$).
In practice, users can measure only the sum of the propagation and queueing delay. and extrapolate a precise measure of the queueing delay from the total delay is highly challenging.

In summary loss-based congestion control algorithms neglect the delay aspect in their algorithms, while delay-based congestion control algorithms always adapt to the delay measure, loosing in terms of performance when competing with loss-based algorithms.
In the next section, we describe our congestion control algorithm which is able to overcome the aforementioned limitations of classical congestion control schemes.

\section{Delay-Constrained Congestion Control Algorithm}\label{ctrl}
\subsection{Control Algorithm}
In this section, we present our DCCC algorithm, showing how it overcomes the main limitations of the existing controllers.
The rate update equation that we consider for our algorithm is the following:
\begin{equation}
	\dot{x}_r = \   k_r x_r \left( U'({x_r}) - v_r( e_r ) - \dot{e}_r  - \pi_r \right)_{x_r}^+  \label{dyna}.
\end{equation}
We now describe the different terms of Eq.~\eqref{dyna} in detail. The parameter $k_r$ tunes the update speed of rate $x_r$. The first term in the brackets is the derivative of the utility function $U'_r(\cdot)$.
The term $v_r(\cdot)$ is the delay penalty function that maps the OWD into a penalty. Similarly to the loss price definition in Eq.~\eqref{eq:loss_ratio}, we write the delay penalty as:
\begin{equation}
	v_r(e_r)=\beta \left( \frac{e_r-T_r}{\text{RTT}_r}\right)^+_{(e_r-T_r)}=\beta \left( \frac{e_r-T_r}{e_r + e_r^b}\right)^+_{(e_r-T_r)},
	\label{del_pen}
\end{equation}
where $\text{RTT}_r$ is the round trip time of user $r$, $e_r$ and $e_r^b$ are the forward and backward experienced delays of user $r$ respectively, $\beta$ is a scaling factor and $T_r$ is the delay threshold of user $r$. The delay threshold is related to delay that the system experiences at the equilibrium. The value of $v_r(e_r)$ is equal to zero if $e_r-T_r<0$ and equal to $\beta(e_r-T_r)/\text{RTT}_r$ otherwise. The normalization of the price by $\text{RTT}_r$ is motivated by rate fairness improvements and stability conditions. Note that this function is not a linear function of the experienced delay, $e_r$, but rather a monotonically increasing function of it.

The derivative of the experienced delay, $\dot{e}_r$, does not modify the equilibrium of the system, since the time derivative at the equilibrium point will be zero by definition. However, it improves the controller performance during the transients, since it provides information about the variation rate of the feedback variable.
The last term in Eq.~\eqref{dyna}, $\pi_r$, takes into account the experienced losses, following Eq.~\eqref{eq:loss_ratio2}, so that the algorithm is able to operate in both delay-based and loss-based scenarios.

In the case of no losses ($\pi_r=0$), our controller behaves as a delay-based controller. The experienced delay at the equilibrium is evaluated by setting the time derivatives to zero in Eq.~\eqref{dyna}:
\begin{equation}
	\hat{e}_r=\frac{\text{RTT}_r}{\beta} U'_r(\hat{x}_r) + T_r = \text{RTT}_r \frac{h}{\beta \hat{x}_r} + T_r.
	\label{eq:eq_delay}
\end{equation}
Substituting Eq. \eqref{eq:RTT_2} into Eq. \eqref{eq:eq_delay} we obtain:
\begin{equation}
	\hat{e}_r=\frac{\hat{e}^b_r h/(\beta\hat{x}_r)+T_r}{1-h/(\beta\hat{x}_r)}.
	\label{eq:eq_delay_2}
\end{equation}
From Eq.~\eqref{eq:eq_delay_2}, we observe that: $i)$ $\hat{e}_r$ converges at the equilibrium to larger values than $T_r$, and approaches to $T_r$ for increasing values of $\hat{x}_r$; $ii)$ the rate at equilibrium has to verify the following inequality:
\begin{equation}
	\hat{x}_r > h/\beta.
	\label{equilibr_ineq}
\end{equation}
This means that the delay price $v_r(e_r)$ in our DCCC algorithm never forces the sending rate to be lower than $h/\beta$.

\begin{figure}
\centering
\includegraphics[scale=0.18]{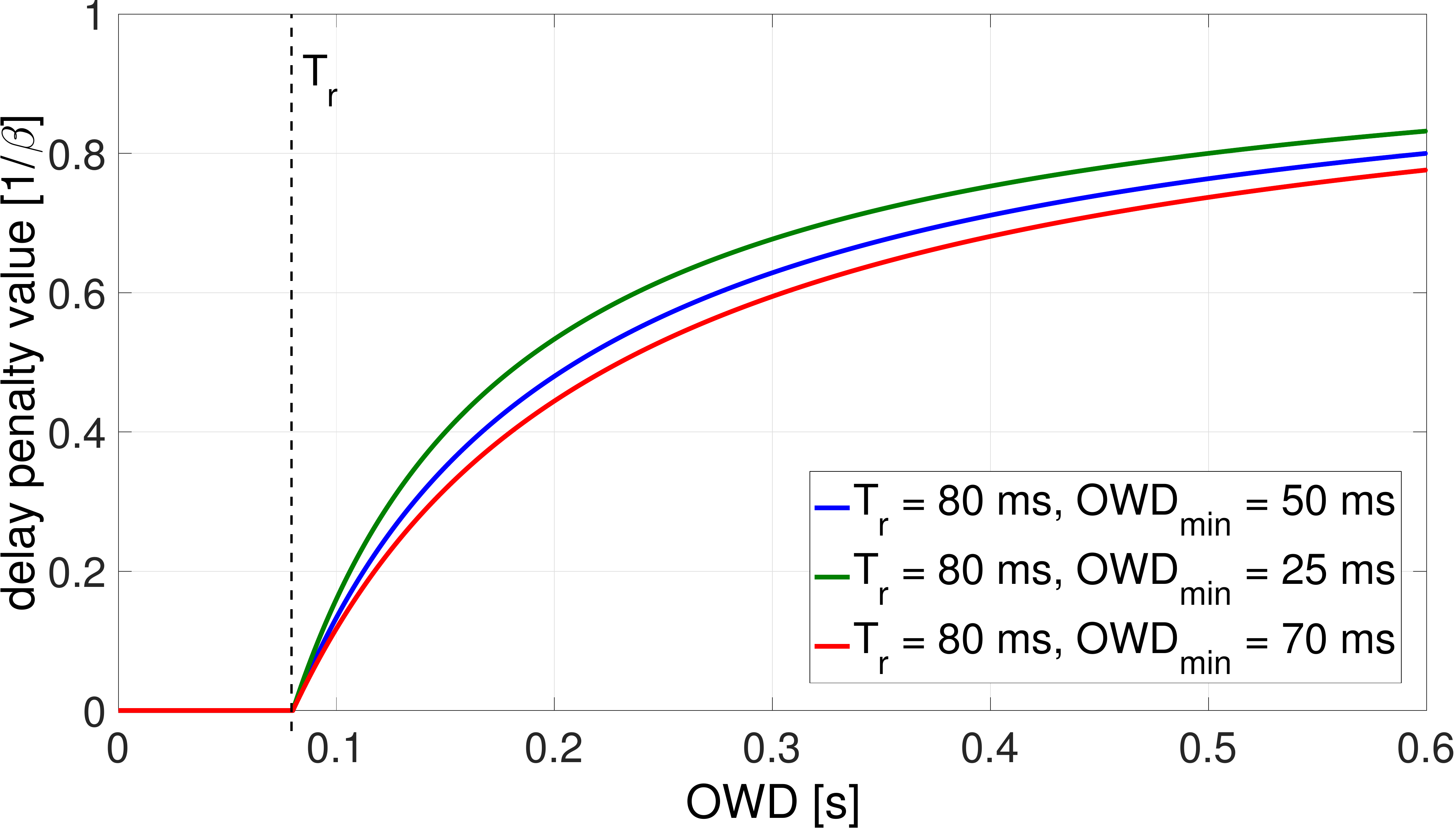}
\caption{Delay penalty as a function of the experienced delay for different values of propagation delays, measured in units of $\beta$.}
\label{penalty}
\end{figure}

The main benefits of our penalty function can be summarized as follows:
\begin{itemize}
\item The non-linearity of the penalty function protects the flows from starvation when competing against loss-based algorithms. Fig. \ref{penalty} depicts the shape of the penalty function of Eq.~\eqref{del_pen} for different values of the propagation delay, when the backwards delay, $e^b_r$ is considered to be equal to the one-way propagation delay in the forward direction. The value of the penalty saturates to $\beta$ for large values of the experienced delay, which is the typical scenario that we encounter when competing with loss-based flows. As a consequence the experienced delay can never force the sending rate to decrease to a value lower than $h/\beta$ thus preventing starvation.
\item The non-linearity of the penalty function alleviates unfairness problems caused by heterogenous propagation delays among the users. Since our control algorithm uses the total experienced one-way delay instead of the queueing delay, it may lead to unfairness when a bottleneck link is shared among users with different propagation delays. However the non-linear mapping of the delay helps to alleviate this problem when the available capacity is low. This can easily be understood by looking at the shape of the penalty function in Fig.~\ref{penalty}. Since the penalty value tends to saturate for large delays, i.e., low available capacity, it means that users with different propagation delays will have similar penalty values in this scenario, and as a consequence similar sending rates.
\end{itemize}

The loss-based part of the algorithm, $\pi_r$, works as an additional penalty that lowers the value of the rate at the equilibrium in case  the delay penalty is not able to reduce it significantly. For instance, when congestion is experienced for low rates, $x_r\leq h/\beta$, the loss ratio $\pi_r$ forces the rate to further decrease. Losses can also happen due to the presence of loss-based flows in the network or to the presence of short buffers inside the network. 
In the extreme case where the maximum end-to-end maximum delay of a route is smaller than the delay threshold, i.e., $e_r^{\text{MAX}}<T_r$, the delay penalty is always zero and the rate is fully driven to the equilibrium exclusively by the loss term $\pi_r$.

\subsection{Controller Implementation}\label{impl}
We now briefly discuss the practical implementation of the DCCC described above.
Our theoretical analysis relates to a continuous system, while in practice we work in a discrete context, where entities that travel through the links are data packets rather than continuous flows. We therefore need to adjust the controller in Eq.~\eqref{dyna} to work in a discrete domain.

In a packet based model a measure that can practically be evaluated is the interarrival time of the packets at the receiver node.
We now explain how, by using this quantity, we are able to build a term that incorporates both aspects of our controller, namely the delay-based part and the loss-based part.
More specifically, the two terms that will be unified are the queueing derivative term, $\dot{e}_r$, and the term $\pi_r$ that takes into account the losses of the flow. We start the derivation of the unified term by finding the connection between the interarrival time and the receiving rate, then we give the expression of the merged term and show that it corresponds to approximate the sum of $\pi_r $ and $\dot{e}_r$

We first consider the case of long buffers where no losses can be experienced. Ideally the interarrival time is inversely proportional to the received rate. If we measure the receiving rate in packets per second, we can write:
\begin{equation}
x_r^{recv} (n)= \frac{1}{t^a(n)-t^a(n-1)}=\frac{1}{\Delta t^a(n)},
\label{estrecv}
\end{equation}
where $t^a(n)$ is the instant of arrival of the $n$-th packet. Noting that $t^a(n)=t^s(n)+e(n)$, where $t^s(n)$ is the sending time of packet $n$, and that the time between two consecutive departures is equal to $1/x_r$, we can write
\begin{equation}
x_r^{recv} (n)= \frac{1}{1/x_r+e_r(n+1)-e_r(n)}=\frac{1}{1/x_r+\Delta e_r(n)}
\label{eq:x_recv}
\end{equation}
where $\Delta e_r(n)$ is the variation of the OWD between two consecutive packets. As the propagation delay is fixed, $\Delta e_r(n)$ corresponds to the variation of the queueing delay. Since the variation is evaluated between the sending time of two consecutive packets, we can write $\dot{e}_r(n) \simeq \Delta e_r(n)x_r $. Recall that we measure the sending rate $x_r$ in packets over seconds, thus $\Delta e_r(n)x_r$ is dimensionless, as $\dot{e}_r$. If we substitute this approximation in Eq.~\eqref{eq:x_recv} and solve for $\dot{e}_r$ we obtain:
\begin{equation}
\dot{e}_r \simeq \frac{x_r-x_r^{recv}}{x_r^{recv}}.
\label{rateprice1}
\end{equation}
The value of $x_r^{recv}$ can be calculated at the receiver using Eq.~\eqref{estrecv} while the value of the true sending rate can be reported in the header of the transmitted packets.

We now analyze the meaning of the right hand side of Eq.~\eqref{rateprice1} in the case of losses. We consider the case of droptail buffers, thus losses are experienced only when the buffers are completely full. The loss ratio for user $r$ is defined as $\pi_r=1-\frac{x_r^{recv}}{x_r}$, or equivalently:
\begin{equation}
\frac{x_r-x_r^{recv}}{x_r^{recv}}=\frac{\pi_r}{1-\pi_r}.
\label{rateprice2}
\end{equation}
Note that the left hand side of Eq.~\eqref{rateprice2} is equal to the right hand side of Eq.~\eqref{rateprice1}. For small loss ratios, $1-\pi_r\simeq 1$, Eq.~\eqref{rateprice2} becomes
\begin{equation}
\frac{x_r-x_r^{recv}}{x_r^{recv}}\simeq{\pi_r}.
\label{rateprice3}
\end{equation}
According to Eq.~\eqref{rateprice3} and Eq.~\eqref{rateprice1} the term $\left(\frac{x_r - x_r^{recv}}{x_r^{recv}}\right)$ can be used in both delay-based scenarios (to approximate $\dot{e}_r$) and loss-based scenarios (to approximate $\pi_r$). The advantage of the merged term is that we do not need to distinguish between the two operation modes of our algorithm, and we simply need to evaluate the value of this term at every rate update step.

We can finally write the rate update rule that governs the DCCC algorithm in the discrete domain as 
\begin{IEEEeqnarray}{rCl} 
x_r^{new}  =  x_r + T_s   k_r  x_r \Big( U'_r({x_r}) - \beta \frac{e_r-T_r}{\text{RTT}_r}  \nonumber \\
- \frac{x_r(t-\text{RTT}_r)-x_r^{recv}}{x_r^{recv}}  \Big),
\label{finalform}
\end{IEEEeqnarray}
where $T_s$ is the time interval at which the rate is updated, and $x_r(t-\text{RTT}_r)$ denotes the delayed sending rate used to calculate the loss and queueing derivative term (the estimation of the received rate and the sending rate need to correspond to the same set of packets for a correct calculation of this final term, thus communication delays cannot be neglected).
In Section~\ref{simulation}, we explain in detail how to effectively set the different parameters of the rate update equation~\eqref{finalform}.

\section{Stability Analysis}\label{stability}
In this section, we discuss the stability of the DCCC algorithm based on Eq.~\eqref{dyna},
focusing on the case of delay-based regime. In order to model a network where the only congestion signal experienced is the OWD, we set the length of network buffers to be infinite. In this case, the flows  experience no losses, (i.e., $\pi_l=0\ \forall l$) and reach the equilibrium point using only the delay measurements. We obviously assume that the equilibrium point exists in this case.
Note that even if the stability of similar delay-based controllers has already been  studied in other works \cite{paganini,stabfasttcp}, due to the non-linear mapping of the experienced delay in Eq.~\eqref{dyna}, those proofs do not apply directly to our algorithm.
The stability of the algorithm in loss-based regime is commented at the end of the section.

We first show the global stability of the non-linear DCCC for an ideal scenario when the communication delays are negligible, i.e., $\tau_{rl}^f \simeq 0$ $\tau_{rl}^b\simeq0$, for a general case of $R$ users and $L$ links. The communication delays are negligible when they are remarkably smaller than the timing characteristics of the control system.
In the second part of the section, we focus instead on local the stability of the controller when communication delays are non-negligible.

\subsection{Negligible Delay Case}
In order to study the global stability of the undelayed system we extend the analysis carried out in~\cite{passivity} to the case of non-linear price function.

\begin{figure}
\centering
\includegraphics[scale=0.24]{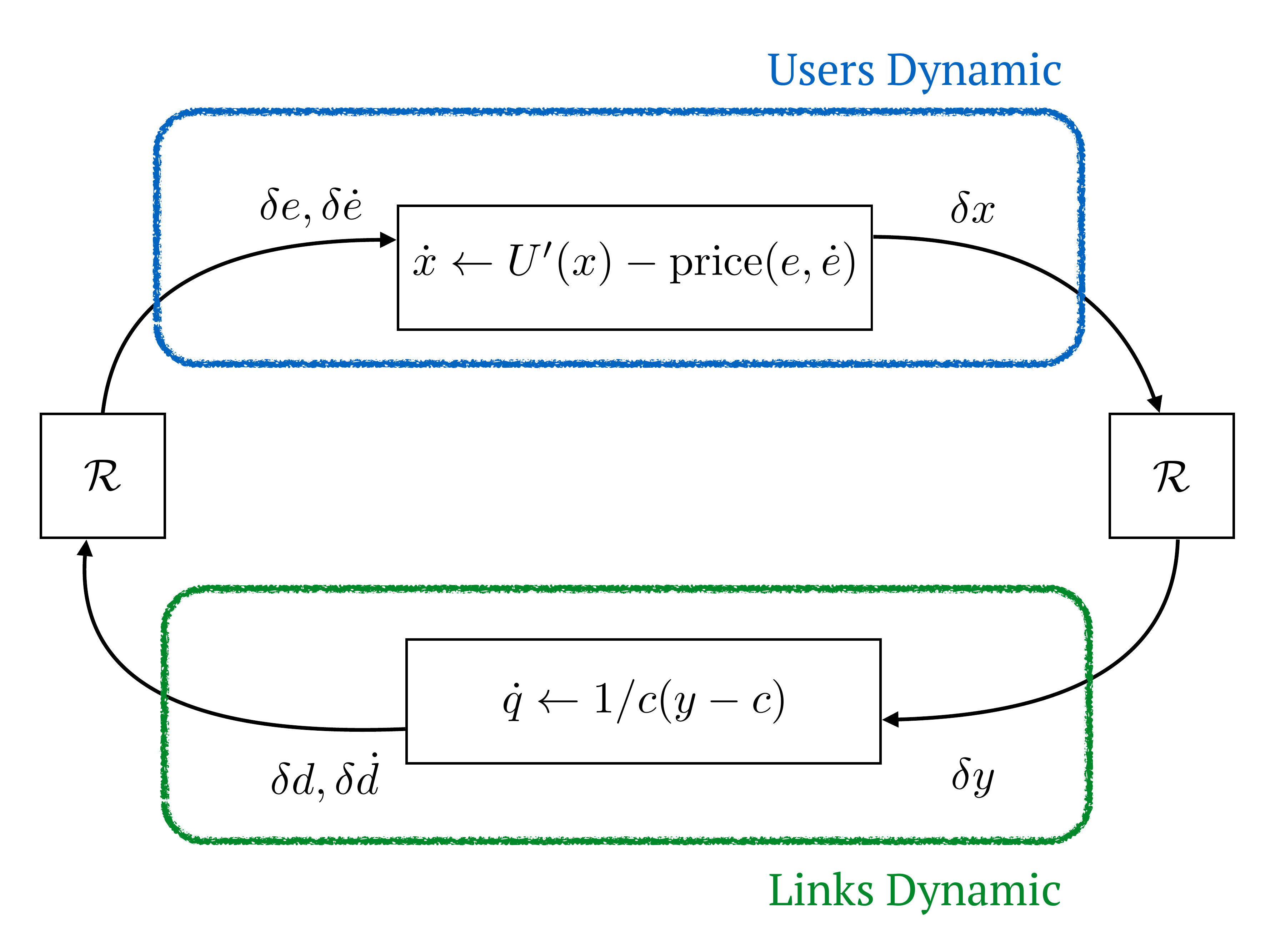}
\caption{Block scheme of the congestion control}
\label{loop}
\end{figure}

The global stability proof is based on the passivity analysis of dynamic systems~\cite{passivity}\cite{passbook}. A dynamic system characterized by an input $u$, state $x$, and output $y$ is said to be passive if there exists a positive definite storage function $V$ such that its derivative can be written as:
\begin{equation}
\dot{V} \leq - W(x) + u^Ty,
\label{eq:pass}
\end{equation}
where $W(x)$ is a positive semi-definite function. If two passive systems are interconnected in a negative feedback loop, then the sum of their respective storage functions is a Lyapunov function for the interconnected system. Indeed, the output of one system is the opposite of the input of the other one, so the two terms $u^Ty$ vanish:
\begin{IEEEeqnarray}{rCl}
\dot{V} &  \leq &- W_1(x_1)- W_2(x_2) + u_1^Ty_1 + u_2^Ty_2 \IEEEyessubnumber \label{sto_2}\\
	   & = & - W_1(x_1)- W_2(x_2) - y_2^Ty_1 + y_1^Ty_2 \IEEEyessubnumber\\
	   & = & - W_1(x_1)- W_2(x_2). \IEEEyessubnumber \label{sto_3}
\end{IEEEeqnarray}
Our dynamic system can actually be modeled as an interconnection of two systems (namely, the users and links dynamics in Fig.~\ref{loop}),
which drive the behavior of our congestion control framework. By proving that these two separate systems are both passive, then we also prove the global stability in the sense of Lyapunov for the entire system.

Let first introduce some notations that are used in our stability proof. The operator $\hat{\cdot}$ denotes the value of the variables at the equilibrium point, and $\delta$ denotes the deviation of a variable from its equilibrium point, e.g.,  $\delta x = x - \hat{x}$. We denote with $V_{user}$ and $V_{link}$ the storage function of the users and links sub-systems respectively. The inputs of the \emph{users} system are the experienced delays $e$ and their derivatives $\dot{e}$, while its output corresponds to the sending rates $x$. Similarly, the \emph{links} system has the link rates, $y$, as input variables and the links delays, $d$ and $\dot{d}$, as output variables. The routing matrix $\mathcal{R}$ maps the user rates to the link rates and the link delays to the users' experienced delays. Finally, the equilibrium point of the system is reached when the dynamic equations \eqref{eq:link_q} and \eqref{dyna} are set to zero. This is equivalent to having $v_r(\hat{e}_r)=U'(\hat{x}_r)$ and $\hat{y}_l=c_l$ if $\hat{q}_l>0$, or $\hat{y}_l\leq c_l$ if $\hat{q}_l=0$. Since we are at equilibrium, $\dot{q}_l=0$; we further have $\dot{e}_r=\sum_{l \in L} \mathcal{R}_{lr} \dot{d}_l=0$, where $d_l$ is the sum of the queueing delay $q_l$ and the propagation delay $p_l$ (which is constant over time).

We consider the following candidate storage functions for the user and link subpart respectively:
\begin{equation}
	V_{user}=\frac{1}{k \alpha} \left( \sum_{r \in R} \int_{\hat{x}_r}^{x_r} \frac{z-\hat{x}_r}{z}dz \right) \label{storage1}
\end{equation}
\begin{equation}
	V_{link}=\frac{1}{2}\sum_{l \in L}  \delta d_l^2 c_l + \frac{1}{\alpha}\sum _{l \in L} (c_l - \hat{y}_l) d_l,\label{storage2}
\end{equation}
for some $\alpha > 0$.
Note that since $d_l\geq 0$ the two storage functions are positive definite functions for $\delta x$ and $\delta q$ respectively. It can be verified, by taking the time derivative of the two storage functions, that inequality of Eq.~\eqref{sto_3} holds for two valid functions $W_{user}$ and $W_{link}$, proving the stability of the system. The full proof can be found in Appendix~\ref{app:proof}.

\subsection{Non-negligible Delay Case}\label{secdelay}
We now study the local stability of the system when the communication between network nodes is affected by non negligible delays, as it is typically the case in practice. In this case we first linearize the rate update equations of the non-linear delayed system at the equilibrium point, then we study the local stability of the delayed system similarly to~\cite{voice,paganini}.

When communication delays are non-negligible, Eq. \eqref{dyna} can be written as:
\begin{IEEEeqnarray}{rCl} 
\dot{x}_r(t) &=& k_r x_r(t) \Big( U'(x_r(t)) -  \label{eq:dynadelay}\\ 
&&  v_r(\sum_{l \in L} \mathcal{R}_{lr} d_l(t-\tau_{rl}^b)) - \sum_{l \in L} \mathcal{R}_{lr} \dot{d}_l(t-\tau_{rl}^b) \Big)_{x_r}^+ \nonumber
\end{IEEEeqnarray}
while the queueing delay given by Eq.~\eqref{eq:link_q} becomes:
\begin{equation}
\dot{q_l}(t) = \frac{1}{c_l} \left( \sum_{r \in R} \mathcal{R}_{lr} x_r(t-\tau_{rl}^f) - c_l \right)_{q_l}^+.
\label{eq:dynadelay2}
\end{equation} 
The new equations \eqref{eq:dynadelay} and \eqref{eq:dynadelay2} represent the users and links dynamics taking into account the communication delays.
Recall that $\tau_{rl}^f$ and $\tau_{rl}^b$ capture the time needed for the packets to go from source $r$ to link $l$, and for the feedback information to be received by the source respectively. Recall also that the congestion information from link $l$ needs to reach the end node of user $r$ before being sent back to the source node. As a consequence, the sum $\tau_{rl}^f+\tau_{rl}^b$ does not depend on $l$ and is equal to the RTT of the route, i.e., $\tau_{rl}^f+\tau_{rl}^b=\text{RTT}_r$.
When we linearize Eq.~\eqref{eq:dynadelay} and Eq.~\eqref{eq:dynadelay2} at the equilibrium point, we obtain the following equations:
\begin{IEEEeqnarray}{rCl} 
\dot{\delta x}(t) &=& k_r \hat{x}_r \Big( \hat{U_r}''(\hat{x}_r) \delta x(t) -  \hat{v}'(\hat{e}_r) \sum_{l \in L} \mathcal{R}_{lr} \delta q_l(t-\tau_{rl}^b)) \nonumber \\
 &-&  \sum_{l \in L} \mathcal{R}_{lr} \delta \dot{q}_l(t-\tau_{rl}^b) \Big), \nonumber \\
\dot{\delta q_l}(t) & = & \frac{1}{c_l} \left( \sum_{r \in R} \mathcal{R}_{lr} \delta x(t-\tau_{rl}^f) \right),
\end{IEEEeqnarray}
where $U_r''$ corresponds to the second derivative of the utility function calculated at the equilibrium point, and $v_r'$ is the derivative of the penalty function calculated at the equilibrium point.
By taking the Laplace transform of the above equation we obtain:
\begin{IEEEeqnarray}{rCl} 
s{\delta x}(s) & = & \mathcal{K} \hat{\mathcal{X}} \left( \hat{\mathcal{U}}'' \delta x(s)- \hat{\mathcal{V}}' \mathcal{R}_b^T(s) \delta q(s) - s \mathcal{R}_b^T(s) \delta q(s) \right)\nonumber \\
s{\delta q}(s) & =  &\mathcal{C}^{-1} \left( \mathcal{R}_f(s) \delta x(s) \right),
\label{linlaplace2}
\end{IEEEeqnarray}
where $\delta x$ indicates the deviation of the variable $x$ from the equilibrium point, i.e., $\delta x =x-\hat{x}$, and $\delta x(s)$ refers to the Laplace transform of $\delta x$. The same notation applies to $\delta q$.
The system of equations \eqref{linlaplace2} is written in a matrix form, where $\mathcal{C}$ is an  $L\times L$ square diagonal matrix whose entries are equal to $c_l$. The routing matrices $\mathcal{R}_f(s)$ and $\mathcal{R}_b(s)$ embed the delay information and are defined as: $\mathcal{R}_{f\ lr}(s)=e^{-s\tau^f_{rl}}$ if user $r$ employs link $l$ and $0$ otherwise, and analogously $\mathcal{R}_{b\ lr}(s)=e^{-s\tau^b_{rl}}$ if user $r$ employs link $l$ and $0$ otherwise. Then, $\mathcal{K}$ and $\hat{\mathcal{X}}$ are $R\times R$ squared diagonal matrices which contain the values of the gains $k_r$ and the values of the rates at equilibrium $x_r$ respectively. $\hat{\mathcal{V}}'$ and $\hat{\mathcal{U}}''$ are $R\times R$ squared diagonal matrices whose entries correspond to the values at the equilibrium point of the first derivative of the penalty delay functions and to the second order derivative of the utility functions respectively.

Next, solving \eqref{linlaplace2} for $\delta x(s)$  we obtain:
\begin{IEEEeqnarray}{rCl} 
{\delta q}(s) & = & \mathcal{C}^{-1} \mathcal{R}_f(s) s^{-1}(Is - \mathcal{K} \hat{\mathcal{X}} \hat{\mathcal{U}}'')^{-1} \nonumber \\
& &\ \ \ \ \ \ \mathcal{K} \hat{\mathcal{X}} \left(- \hat{\mathcal{V}}' - s \right) \mathcal{R}_b^T(s) \delta q(s).  \label{pre_loop}
\end{IEEEeqnarray}
From Eq. \eqref{pre_loop} we can easily determine the return loop ratio $F(s)$~\cite{macfarlane,kellyvoice}, of our multivariable feedback system:
\begin{equation}
F(s) = \mathcal{C}^{-1} \mathcal{R}_f(s) s^{-1}(Is - \mathcal{K} \hat{\mathcal{X}} \hat{\mathcal{U}}'')^{-1} \mathcal{K} \hat{\mathcal{X}} \left(\hat{\mathcal{V}}' + s \right) \mathcal{R}_b^T(s).
\end{equation}
Noting that $ \mathcal{R}_b^T(s)=\text{diag}({e^{-s\text{RTT}}})\mathcal{R}_f^T(-s)$ and using the property that diagonal matrices commute, we can write:
\begin{IEEEeqnarray}{rCl} 
F(s) & = & \mathcal{C}^{-1} \mathcal{R}_f(s) s^{-1}(Is - \mathcal{K} \hat{\mathcal{X}} \hat{\mathcal{U}}'')^{-1} \nonumber \\
&& \ \ \ \ \mathcal{K}  \left(\hat{\mathcal{V}}' + s \right) \text{diag}({e^{-s\text{RTT}}}) \hat{\mathcal{X}} \mathcal{R}_f^T(-s). \label{loopR}
\end{IEEEeqnarray}
For the sake of clarity, we introduce the matrix $G(s)$:
\begin{equation}
G(s) = s^{-1}( Is - \mathcal{K} \hat{\mathcal{X}} \hat{\mathcal{U}}'')^{-1} \mathcal{K}  \left(\hat{\mathcal{V}}' + s \right) \text{diag}({e^{-s\text{RTT}}}), \label{G_eq}
\end{equation}
this is an $R\times R$ diagonal matrix. Similarly to~\cite{stabfasttcp}, we further introduce the matrix
\begin{equation}
 {\tilde{\mathcal{R}}}(s)=\text{diag}(1/\sqrt{c_l})\mathcal{R}^f(s)\text{diag}(\sqrt{x_r}).
\end{equation}
Since the eigenvalues of matrix product does not depend on the order of the matrices we can write: 
\begin{equation}
\sigma(F(s)) = \sigma \left( \tilde{\mathcal{R}}(s) G(s) \tilde{\mathcal{R}}^T(-s) \right),
\label{eq:F}
\end{equation}
where $\sigma(\cdot)$ denotes the set of eigenvalues of a matrix.
We can now verify the stability of the system using the Nyquist stability criterion. We set $s=j\omega$, where $j$ is the imaginary unit, and we vary its value on the Nyquist path. The trajectories of the eigenvalues of the system as a function of $j\omega$ must not encircle the point $-1+j0$ for the system to be stable. We use now the main result of \cite{vinnicombe}:
\begin{equation}
\sigma(F(j\omega)) \in \rho (\tilde{\mathcal{R}}^T(j\omega)\tilde{\mathcal{R}}(-j\omega))\ \text{co}(0 \cup \sigma(G(j\omega))),
\label{convex}
\end{equation}
where $\rho({\cdot})$ denotes to the spectral radius of a matrix and $\text{co}(\cdot)$ denotes the convex hull. Eq. \eqref{convex} states that the eigenvalues of $F(j\omega)$ are located on the convex hull made by the eigenvalues of $G(j\omega)$ scaled by the spectral radius of $\tilde{\mathcal{R}}^T(j\omega)\tilde{\mathcal{R}}(-j\omega)$.

From \cite{vegas}, we further know that the spectral radius $\rho (\tilde{\mathcal{R}}^T(j\omega)\tilde{\mathcal{R}}(-j\omega)) \leq M$, with $M$ being the maximum number of links used by any user $r$. 
Hence, to prove the stability of the system, we need to show that the eigenvalues of $G(j\omega)$ in Eq.~\eqref{eq:F} do not encircle the point $-1/M+j0$ in the complex plane. Since $G(j\omega)$ is a diagonal matrix, as can be seen from Eq.~\eqref{G_eq}, we need to show that all the entries on its diagonal verify the Nyquist stability criterion. Defining $G_r(j\omega)$ as the $r$ element of the diagonal of matrix $G(j\omega)$ we have
\begin{equation}
G_r(j\omega)=k_r\frac{e^{-j\omega \text{RTT}_r}}{j\omega}\frac{v'_r+j\omega}{j\omega-k_r \hat{x}_rU_r''}.
\label{eq:G_r}
\end{equation} 
As the utility functions are concave, $U_r''$ is negative, and so both poles of Eq. \eqref{eq:G_r} are non positive. With a utility function defined as $U_r(x_r)=h\log(x_r)$, the value of the second derivative of the utility and the derivative of the price at the equilibrium point are respectively equal to
\begin{equation}
U_r''=-\frac{h}{\hat{x}_r^2}\ \ \ \ \  v_r'=\frac{\beta-h/\hat{x}_r}{\text{RTT}_r},
\label{eq:v_der_U_der}
\end{equation}
where in the second equation we used the property that $v_r(\hat{e}_r)=U'(\hat{x}_r)$ at equilibrium. Note that, in order to have a negative zero in the transfer function $G_r(j\omega)$ we need the inequality $\hat{x}_r>h/\beta$ to hold. This condition supports in fact that $h/\beta$ is the minimum equilibrium rate for the purely delay-based subpart as described in Section~\ref{ctrl}.

Finally, substituting Eq.~\eqref{eq:v_der_U_der} in Eq.~\eqref{eq:G_r}, we obtain
\begin{equation}
G_r(j\omega) = \eta_r \frac{e^{-j\omega \text{RTT}_r}}{j\omega \text{RTT}_r}\frac{\frac{\beta-h/\hat{x}_r}{\text{RTT}_r}+j\omega}{j\omega+\eta_r\frac{h}{\hat{x}_r\text{RTT}_r}},
\label{TF_2}
\end{equation}
where we have introduced the  normalized gain $\eta_r=k_r\text{RTT}_r$.
Eq. \eqref{TF_2} is a transfer function with two poles and one zero. The location of the second pole and the zero are actually linked since they depend on the same quantities. In particular, due to the previous considerations about the minimum value of the equilibrium rate, the pole cannot be located at an angular frequency larger than $\eta_r\beta/\text{RTT}_r$, and the zero has to be located between 0 and  $\beta/\text{RTT}_r$.
By selecting appropriate values for the controller parameters $h$, $\beta$ and $\eta_r$, we can ensure that point $-1/M+j0$ is not encircled. The tuning of the parameters is complex because in general the setting of one parameter affects the value of the others. One possible way to proceed is to allocate the zero-pole couple at low frequencies and the cross frequency at which $|G_r(j\omega)|=1/M$ to a higher value.  In this case at the cross frequency the zero and the pole compensate each other decoupling and facilitating the tuning of the rest of the parameters.

\subsection{Summary of stability analysis}
In this section we have studied  the stability of our system in the case where the only congestion event received from the network is the experienced delay. The main result is that the non-linear mapping of the experienced delay leads to a stable  delay-based system. 
The stability analysis can also be extended to lossy scenarios.
In the case of droptail buffer policy, when losses are experienced  the queue size is constant  and equal to the maximum one (we do not consider the case of Random Early Detection (RED) policy~\cite{RED} or any other Advanced Queue Management (AQM) mechanism). Since a constant delay plays no role in the placement of eigenvalues of the linear dynamic system at equilibrium, the delay-based terms of the rate update equation disappear in the linearization process of the system. Hence, for the loss-based part of our algorithm, the linearized update equation is consistent with the one used in~\cite{kelly,massoulie} and the stability proof can be carried out by following the steps described in these works.

\section{Final Implementation and Simulations}\label{simulation}
We now provide simulations results for the proposed DCCC algorithm. First, we explain how the controller performance is affected by each parameter and how to efficiently set them. Then we provide the simulation results where we analyze the basic behavior of our algorithm, the intra-protocol fairness, the TCP coexistence and finally we provide a comparison with other similar congestion control algorithms.

\subsection{Parameters and Simulation Setup}
The parameters of the DCCC are set by using the results obtained from the stability analysis of Subsection~\ref{secdelay}, and by empirical knowledge on the effect of the parameters on the controller behavior.

The gain $k_r$, the sampling period $T_s$ and the parameter $\beta$ of Eq.~\eqref{finalform} strongly affect the cross frequencies of the transfer functions that compose the matrix $G$ (defined in Eq.~\eqref{G_eq}). They therefore affect the speed and the stability of the closed loop system and are set to impose the stability of the system.

Conversely, the parameters $h$ and $T_r$ do not strongly affect the stability of the system. 
The value of $h$ affects the aggressiveness of the controller, as the price at equilibrium is proportional to $h$. The price reflects both the overshoot of the delay threshold in delay-based environments and the fairness level with other loss-based protocols. 
The value of $h$ reflects also the minimum rate that is achieved with extremely long buffers by the delay penalty, which is equal to $h/\beta$.
Therefore the value of $h$ has to be properly tuned to find the best tradeoff between large guaranteed rates (i.e., large $h$) and not too large prices (i.e., small $h$).

The delay threshold  $T_r$ is set by the application. Ideally, if the two endpoints are synchronized, the application should set $T_r$ such that an acceptable end-to-end delay is achieved. It is worth noting that in order to guarantee a fully utilization of the channel, $T_r$ should be larger than the minimum OWD.

In Table~\ref{tab:par} we list for each controller parameter the value adopted in the conducted simulations as well as a suggested operation range. Finally, we point out that the parameters setting that we use are valid in a wide range of scenarios, even if they might not always lead to optimal performance. For example, a $T_r$ value between $50$ ms and $100$ ms represents a valid setting in a wide range of network condition, provided that it is tolerated by the application. For a more thorough discussion on the setting of the parameters, and on how they affect the algorithm behavior, we refer the reader to Appendix~\ref{app:param}.

\begin{table}
\centering
\caption{Paramers setting Summary}
\label{tab:par}
\begin{tabular}{c|c|c}
Parameter name & Used value   & Suggested range       \\ \hline \hline
$k_r$          & $1/(2.5\text{RTT})$ & $1/(2.5T_s)$-$1/(10T_s)$  \\ \hline
$\beta$        & $0.1$        & $0.05$-$0.2$\\ \hline
$T_s$          & RTT          & $50$-$500$ ms\\ \hline
$T_r$          & $50$-$100$ ms & $5$-$250$ ms\\ \hline
$h$            & $20$ kbps  &  $5$-$50$ kbps \\ \hline \hline
\end{tabular}
\end{table}

To evaluate the performance of our controller, we carried out experiments using the NS3 network simulator platform. We have tested the controller in different network topologies and scenarios in order to show that the algorithm is able to work in loss-based as well as delay-based environments. In particular, we consider a single link topology, Topology $1$ (see Fig.~\ref{topo1}) and Topology $2$ (see Fig.~\ref{topo2}) in our simulations. The first one is the classic dumbbell topology, where several users share the same unique bottleneck link. The second topology is the so-called parking lot topology, with two bottleneck links, where two users employ only one of these links while a third user, with a longer data path, employs both congested links. All the links connecting the endpoints to the bottleneck are set to a high connection speed, e.g. $100$ Mbps. We focus on low/medium values of the bottleneck capacity since this is the typical, and most critical, scenario for real-time applications, see~\cite{rmcattestcases}. We evaluate the performance of the DCCC algorithm under different metrics such as throughput, self-inflicted delay and fairness. We also compare the algorithm with other delay-based congestion controls, namely: the Network-Assisted Dynamic Adaptation (NADA) congestion control algorithm~\cite{nada}, the Google congestion control (GCC) algorithm~\cite{google-draft} and the Low Extra Delay Background Traffic (LEDBAT) algorithm~\cite{ledbat}. For a more detailed description of the congestion control algorithms implementation and for further simulation test cases, we refer the reader to Appendix~\ref{app:imp} and Appendix~\ref{app:sim}.

\subsection{Fairness Analysis}
\begin{figure}
\centering
\includegraphics[scale=0.18]{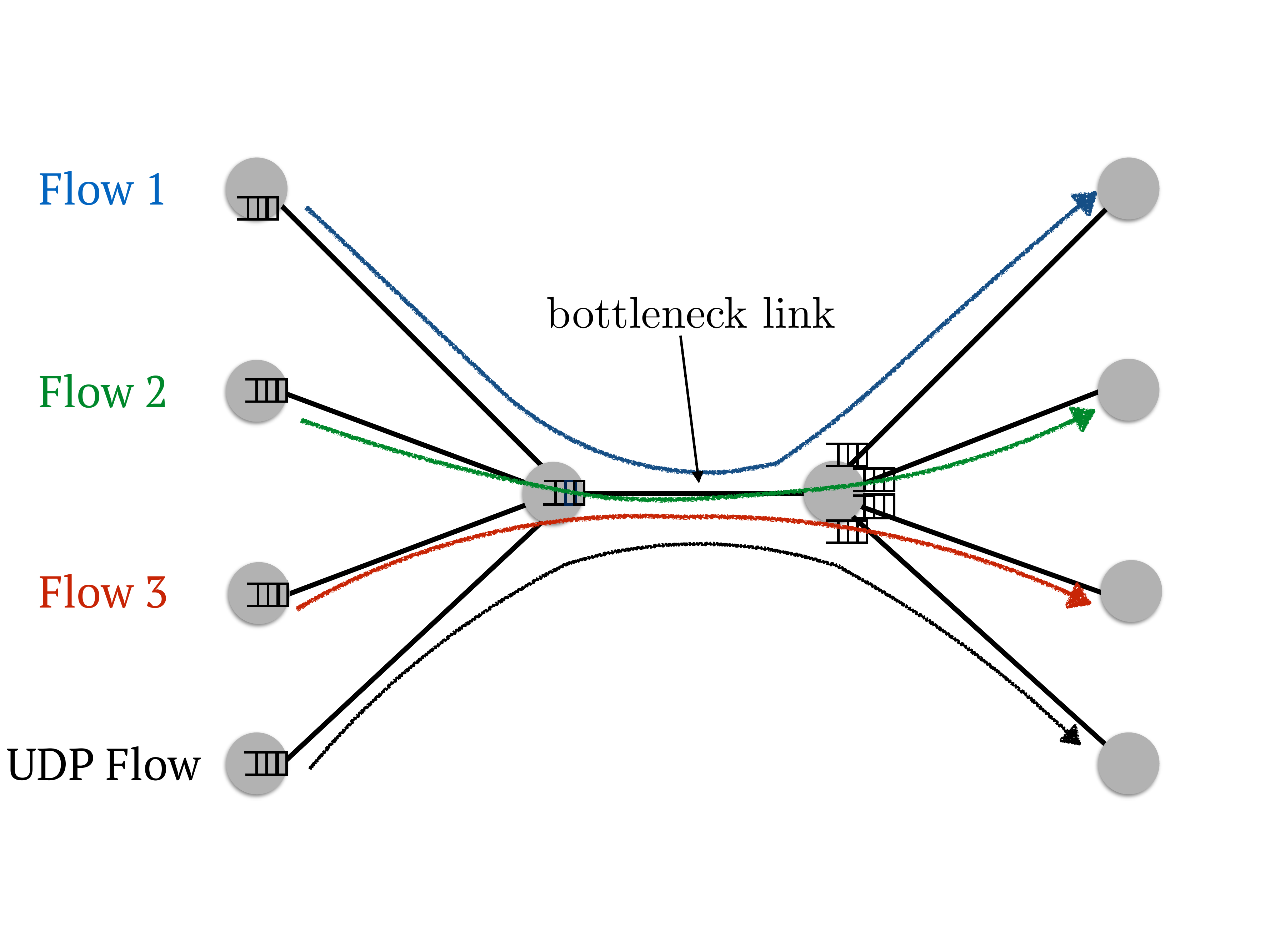}
\caption{Network Topology $1$.}
\label{topo1}
\end{figure}

In this set of simulations, we evaluate the performance of the DCCC controller when it shares the bottleneck links with other flows controlled by the same algorithm. Fairness is an important metric for congestion control algorithms; since it reflects the ability of flows to fairly share the available bandwidth without penalizing early or late starting flows.
We consider a case with three DCCC flows that share a link with an unresponsive UDP flow with a constant rate of $500$ kbps. The network topology is shown in Fig.~\ref{topo1}. Two of the flows ($1$ and $2$) start  at $2$ and $4$ s respectively and last until the end of the simulation. The flow 3 starts at $100$ s and stops transmitting at $260$ s. The unresponsive UDP flow is always active during the simulation. The bottleneck link has a one way propagation delay of $25$ ms, and a total capacity of $3.5$ Mbps. The threshold delay of the three DCCC flows is set to $100$ ms.

\begin{figure}
\centering
\includegraphics[scale=0.1]{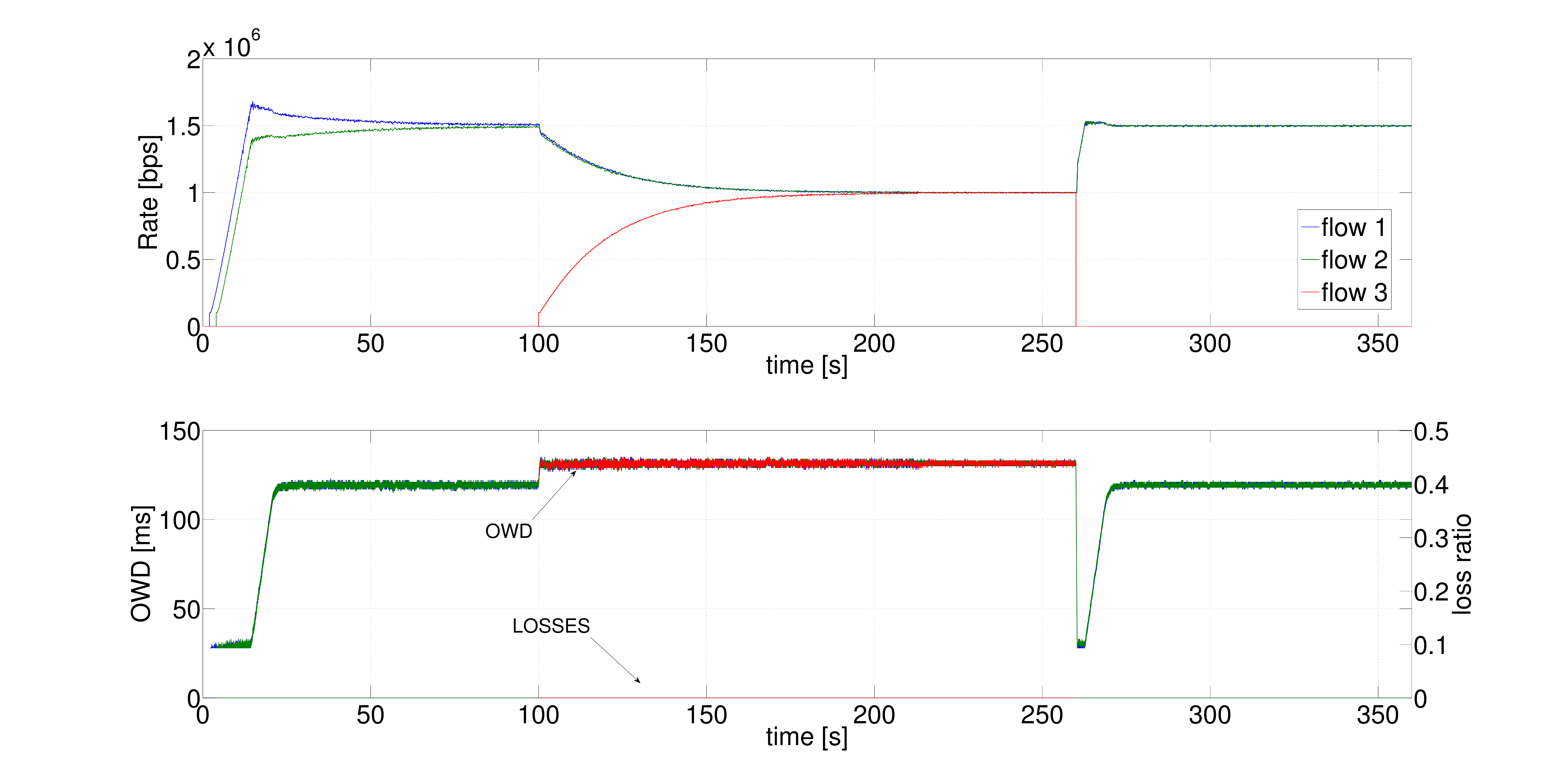}
\caption{Sending rate and OWD for the three DCCC flows sharing a common bottleneck link when the equilibrium is driven by the experienced delay.}
\label{delay}
\end{figure}

In a first simulation we set the maximum buffer length to $130$ packets (approximately $300$ ms of maximum queueing delay). Since $T_r\ll300$ ms, no losses are experienced by the flows at equilibrium. Results are provided in Fig.~\ref{delay}. The DCCC flows fairly share the bottleneck link without penalizing early or late starting flows. When all three flows are active the delay at equilibrium is slightly higher, as expected since the sending rate at the equilibrium goes from $1.5$ Mbps per user to $1$ Mbps per user. This behavior is indeed expected: according to Eq.~\eqref{finalform} the algorithm needs a higher price, i.e., a larger delay, in order to reach a lower rate at equilibrium.

\begin{figure}
\centering
\includegraphics[scale=0.1]{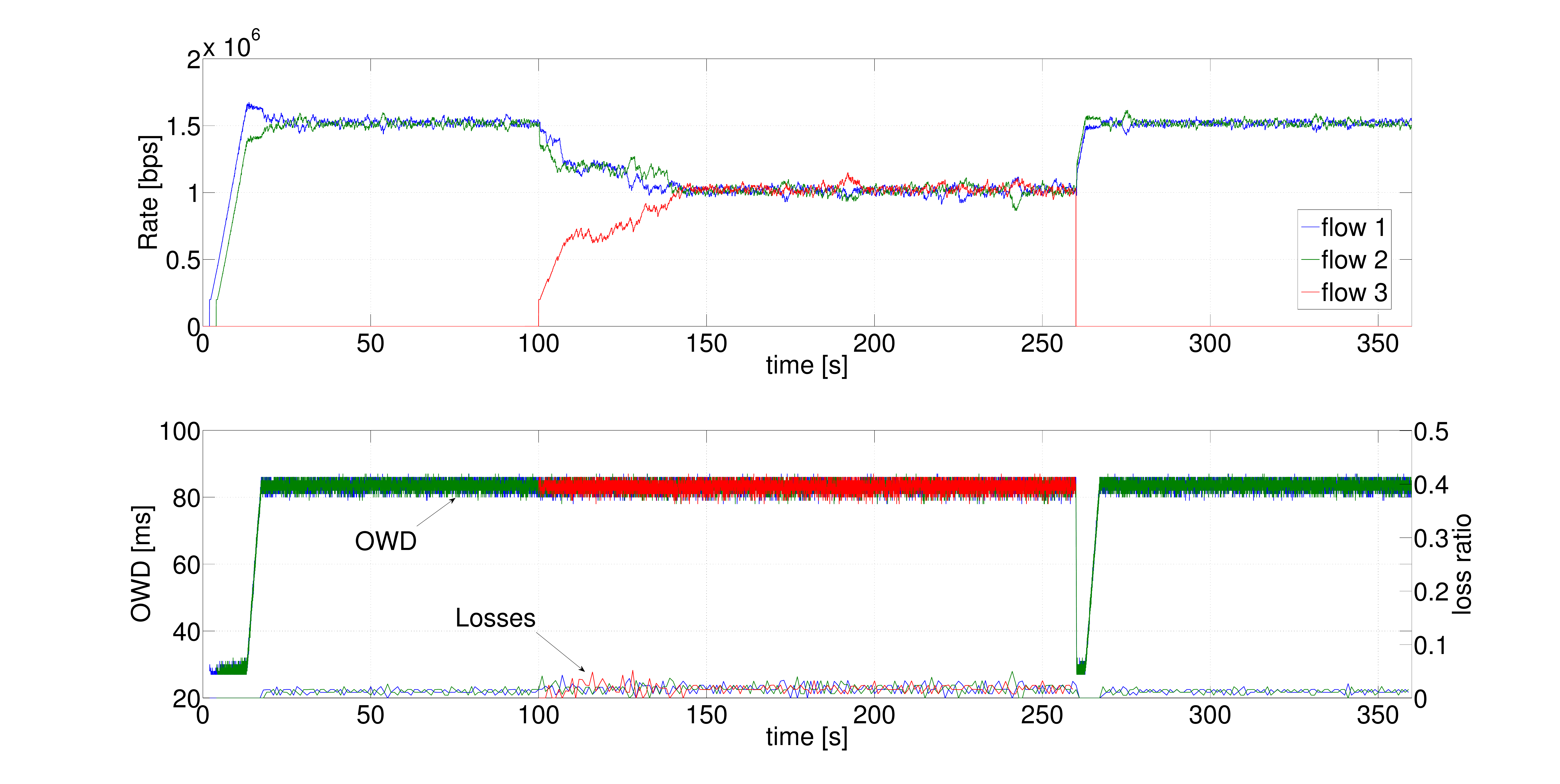}
\caption{Three flows sharing a common bottleneck, equilibrium is driven by losses. Drop tail bottleneck buffer.}
\label{droptail}
\end{figure}

We then conduct a simulation with the same settings, but with a maximum buffer length of $25$ packets (approximately $45$ ms of maximum queueing delay). In this scenario $T_r$ is greater than the maximum experienced delay, therefore the DCCC flows are driven to equilibrium by the experienced losses rather than by the experienced delay. The sending rates and OWD for this simulation are depicted in Fig.~\ref{droptail}. We observe that, also when the system is in a loss-based regime, the flows fairly share the available bandwidth.

\begin{figure}
\centering
\includegraphics[scale=0.18]{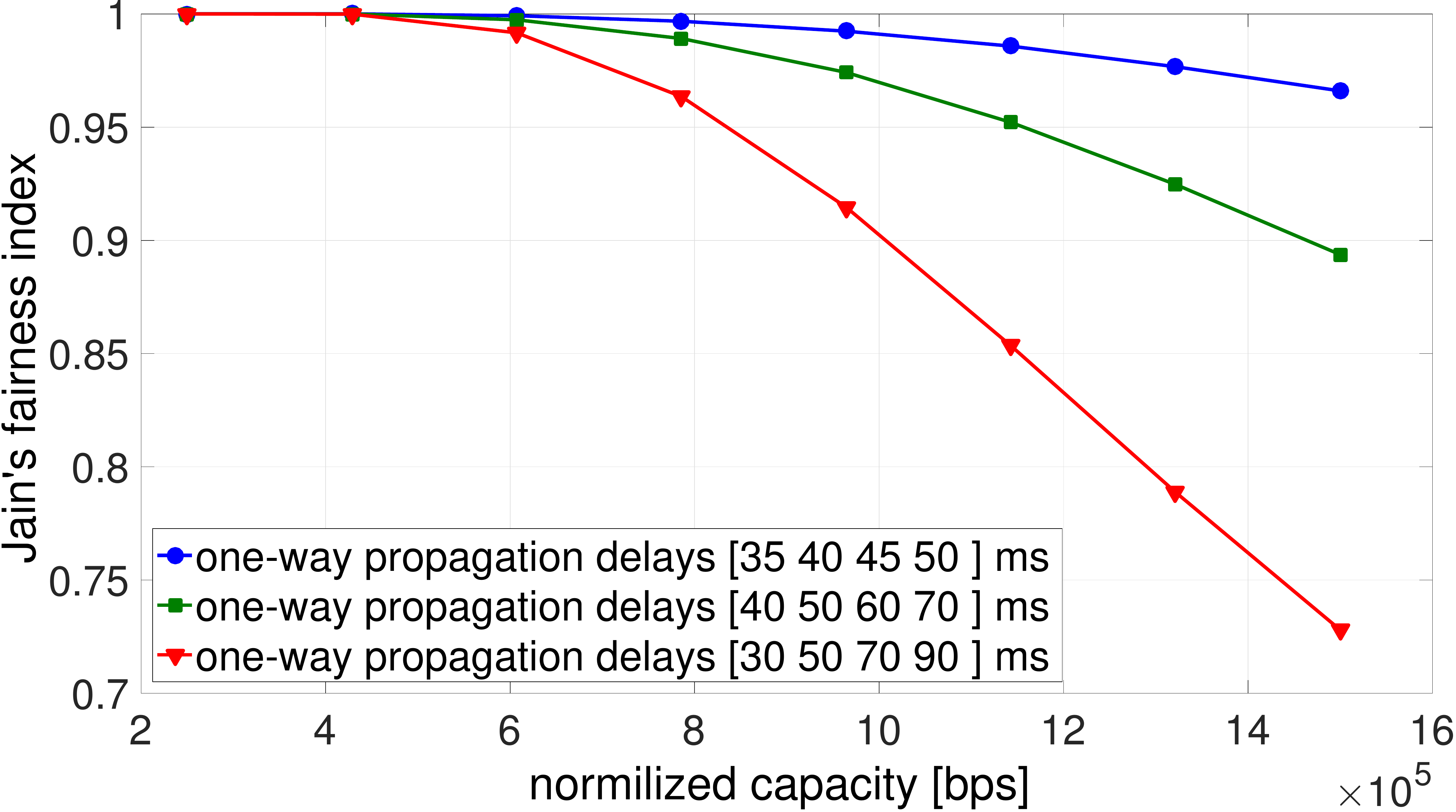}
\caption{Jain's fairness index of four DCCC flows with different propagation delays.}
\label{jain}
\end{figure}

\begin{figure}
\centering
\includegraphics[scale=0.18]{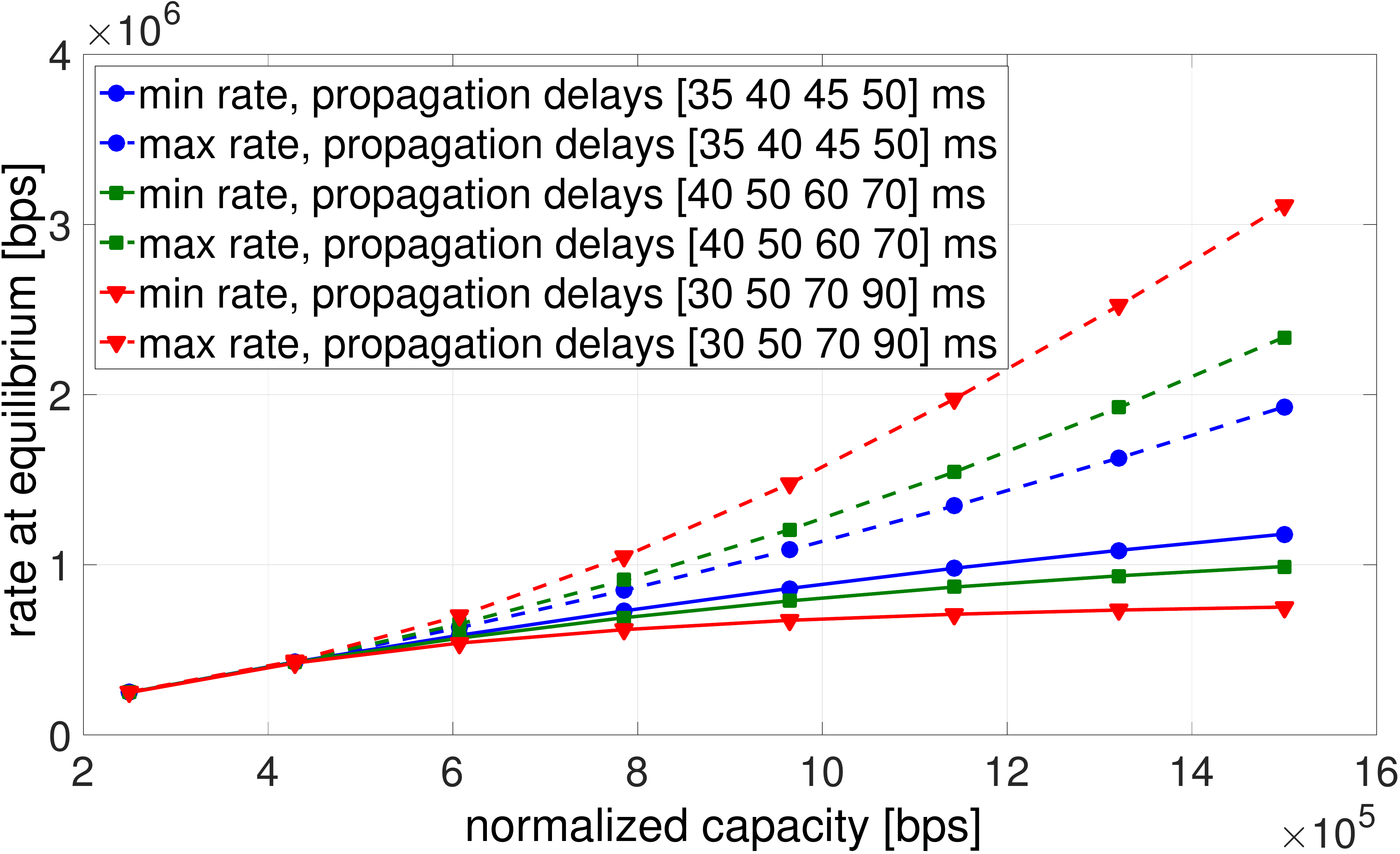}
\caption{Minimum and maximum rates at equilibrium versus normalized capacity for four DCCC flows sharing a bottleneck with different propagation delays.}
\label{fair2}
\end{figure}

As second scenario, we still consider the network topology of the previous simulations, but in this case the flows have different propagation delays, therefore different prices.
We denote by $[p_1\ p_2\ p_3\ p_4]$ the propagation delays of the four considered flows. The capacity of the bottleneck link varies from $1$ Mbps to $6$ Mbps. The maximum queueing delay is set to $500$ ms (the buffer size ranges from about $60$ packets to $360$ packets depending on the capacity value), and $T_r$ is set to $100$ ms. We measure the average sending rate at equilibrium and we compute the Jain's fairness index~\cite{jain}, which for the case of $N$ flows is defined as: 
\begin{equation}
J(x)=\frac{\left( \sum_{n=1}^N x_n \right)^2 }{N \sum_{n=1}^N x_n^2}.
\end{equation}
The Jain's Fairness index is a well known index to measure the fairness of rate allocation and it ranges between $1$ (full fairness) and $1/N$ (poor fairness).
The simulation results are shown in Fig.~\ref{jain}, where the fairness index is provided as a function of the normalized capacity,  which corresponds to the total capacity normalized by the number of flows. The most heterogenous scenario (e.g. OWDs equal to $[30\ 50\ 70\ 90]$) is the least fair especially for large capacity values. However, when the available bandwidth is limited, which is the most constrained and  the most critical scenario, the fairness is improved. This behavior is motivated by the fact that, even if heterogenous propagation delays lead to heterogenous prices, the mismatch is reduced for large prices (small capacities) leading to a more fair rate allocation.
Fig.~\ref{fair2} depicts the users' lowest  and highest sending rate at equilibrium as a function of the normalized capacity. In the ideal case of full fairness, this plot should feature a straight line with all the rates equal to the normalized capacity. 
Even if curves in Fig.~\ref{fair2} deviates from this ideal behavior for high values of capacity, still no flow is experiencing starvation and a relatively good sending rate  is achieved.
For example, when the maximum propagation delay of one flow is three times the minimum one, the most penalized flow is still able to achieve a sending rate equal to approximately half normalized capacity. 
We conclude these results by pointing out that, unlike loss-based algorithms, for delay-based algorithms achieve full intra-protocol fairness in all possible scenarios is particularly challenging. The challenge is caused by the fact that  flows might not commonly agree on the value of the price. Note that this issue is not a peculiarity of our algorithm, for example, in other protocols~\cite{nada,questledbat}, latecomers flows\footnote{A latecomer flow is a flow that starts sending data over an already congested (i.e., standing queueing delay grater than zero) network link.} tend to underestimate the queuing delay and converge to larger sending rates. 
In our algorithm we are not able to guarantee optimal intra protocol fairness in all possible scenarios, but we mitigate unfairness issues in the most constrained scenarios, sacrificing the performance in less constrained scenarios, i.e., when capacities are large.
 
\begin{figure}
\centering
\includegraphics[scale=0.18]{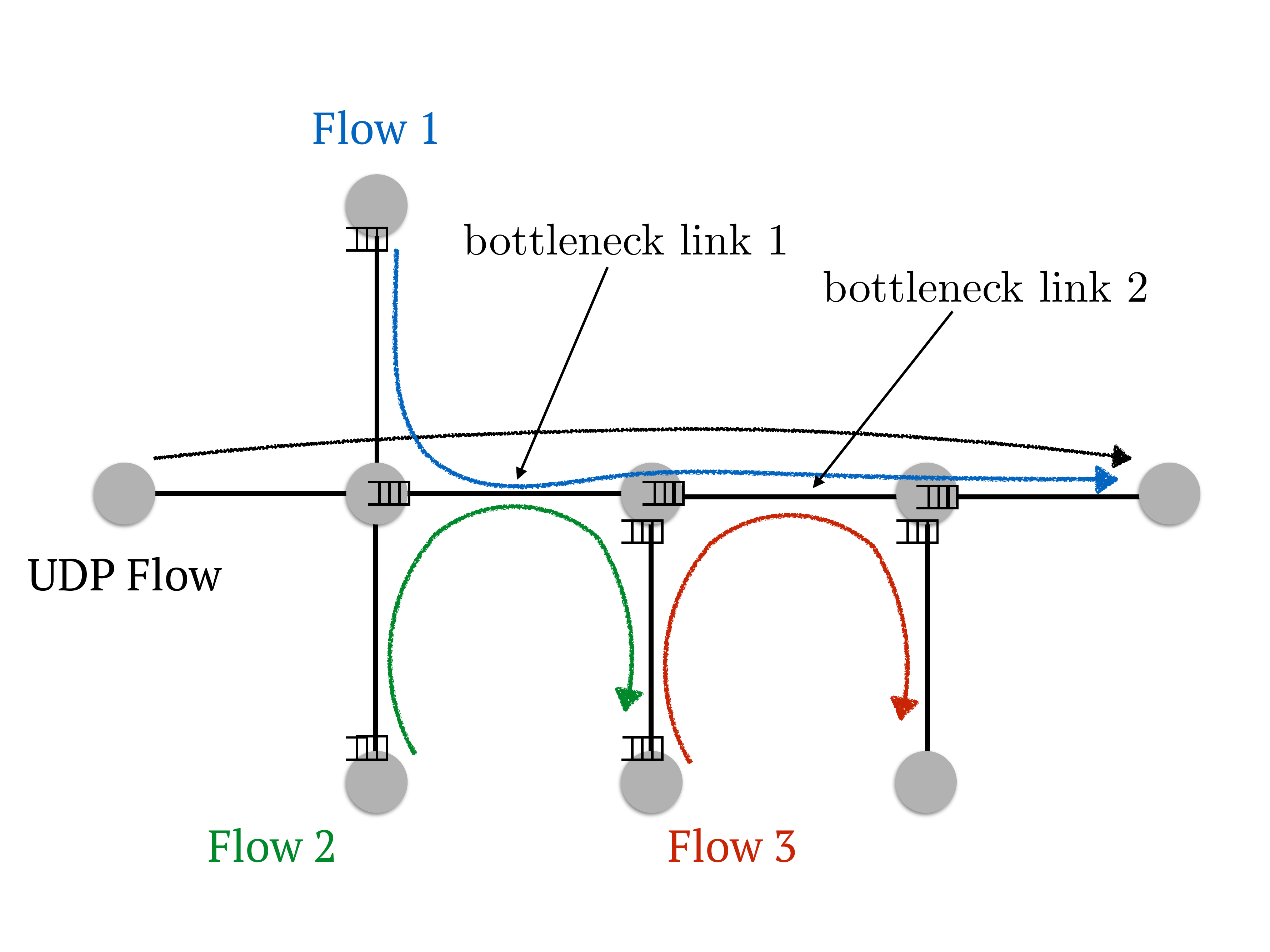}
\caption{Network Topology $2$. The flow DCCC $1$ passes through two bottlenecks links and competes with another DCCC flow on each of these links}
\label{topo2}
\end{figure}

\begin{figure}
\centering
\includegraphics[scale=0.1]{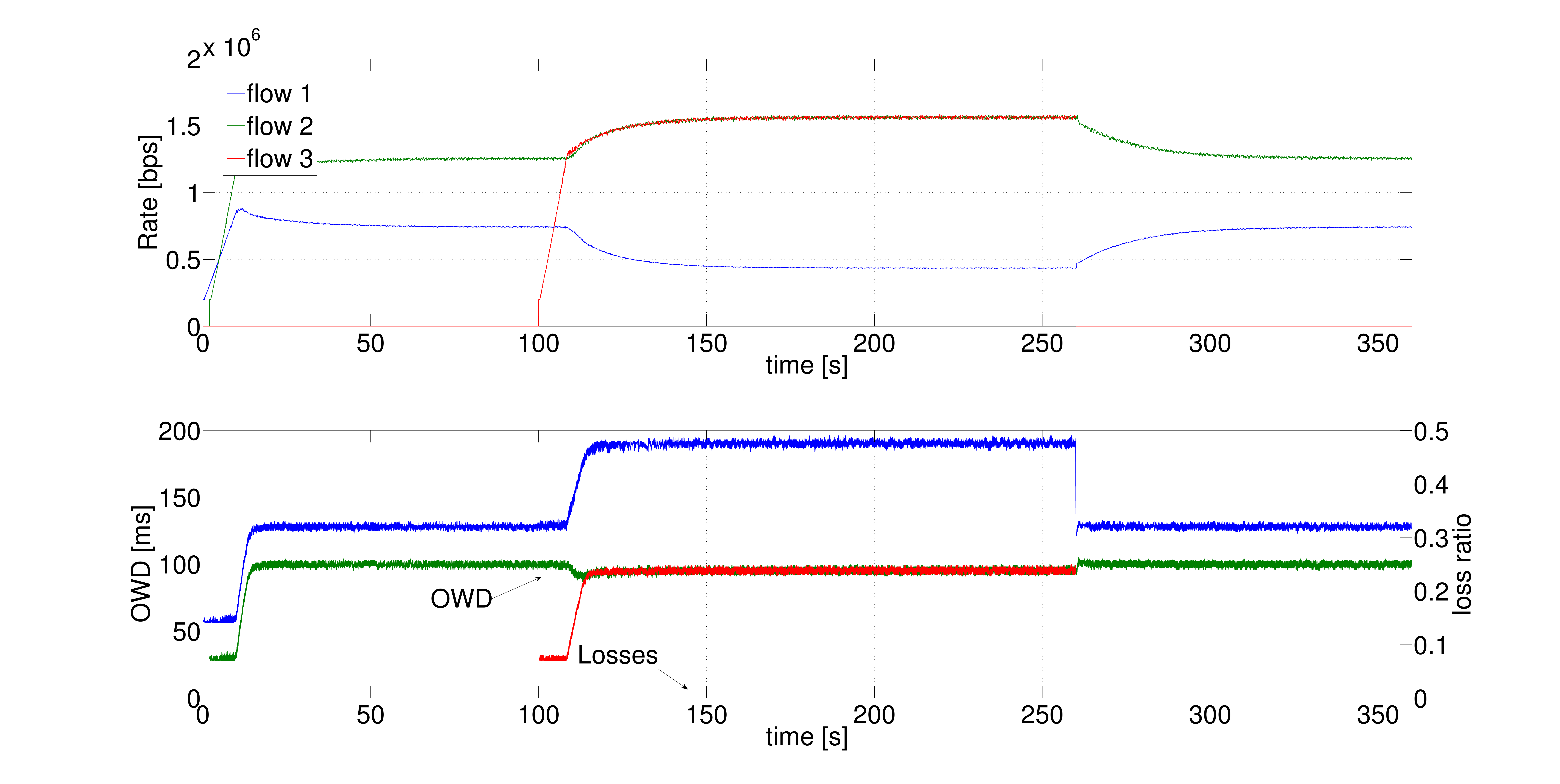}
\caption{Time evolution of the sending rates and delays for the parking lot topology when equilibrium is driven by the experienced delay.}
\label{delay2}
\end{figure}

We now analyze the fairness of our controller for the parking lot network topology depicted in Fig.~\ref{topo2}. 
Both bottleneck links have a channel capacity of $2.5$ Mbps and a propagation delay of $25$ ms. An unresponsive UDP flow streams through both links at a constant rate of $500$ Kbps. For the equilibrium to be driven by the experienced delay, we set the threshold $T_r$ to $100$ ms for all the users, and the maximum length of the buffers inside the network to $100$ packets (equivalent to a maximum queueing delay of approximately $330$ ms), which ensures that no losses are experienced during this simulation. Flow $3$ starts at $100$ s and stops at $260$ s while the other flows are active during the entire simulation. The simulation results are shown in Fig.~\ref{delay2}. Flow $1$  always  gets a lower rate compared to the other flows. This is expected and due to the fact that its route is longer, hence it is penalized compared to the other flows.

\begin{figure}
\centering
\includegraphics[scale=0.1]{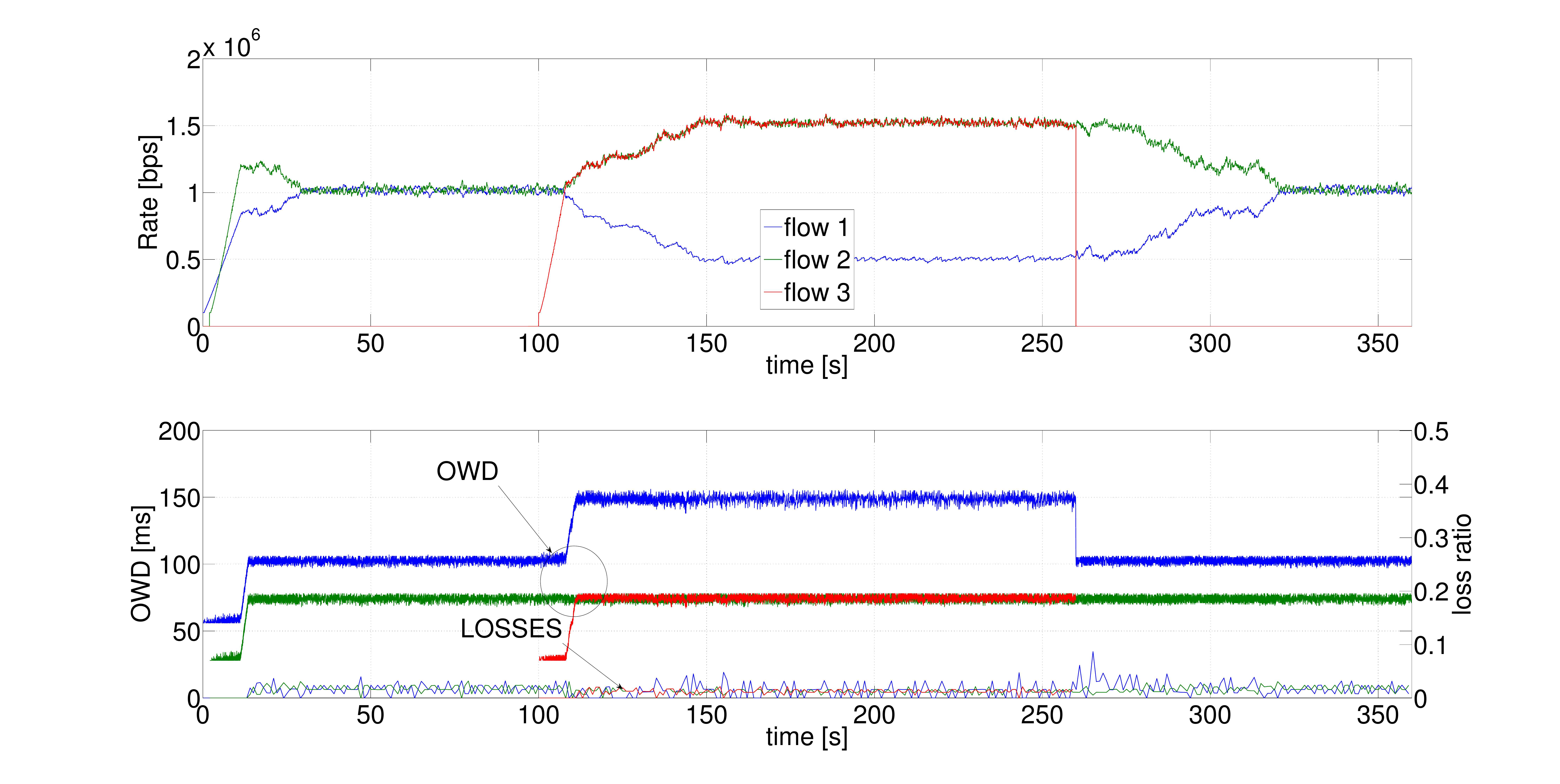}
\caption{Time evolution of the sending rates and delays for the parking lot topology when equilibrium is driven by losses}
\label{losses2}
\end{figure}

Finally, we conduct a last simulation for a mixed regime: the maximum queuing delay is set to $15$ packets (equivalent to a maximum queueing delay of approximately $50$ ms), while the threshold delay $T_r$ is set to $100$ ms. Also in this case we observe that flow $1$ achieves a lower rate than the other two competing flows, as shown in Fig.~\ref{losses2}. In this experiment flow $1$ is experiencing losses and having a positive value of the delay penalty, hence we call this operation mode mixed regime. Note that the maximum queuing delay is set on a per buffer basis, thus flow $3$, which passes through two congested links, experiences a total queuing delay of $100$ ms. In this test case, flows $1$ and $2$ converge to the same rate when user $3$ is not active because both pass through the same unique bottleneck, and the OWD is lower then the threshold delay for both of them. On the other hand, when all three flows are active, flow $1$ traverses two bottlenecks links, experiencing a double value of losses and converging to a lower equilibrium rate.


 \subsection{TCP Coexistence}\label{sec:tcp}
We now study the performance of the DCCC when it competes against TCP flows. We again use a single link topology, with a capacity of $2.5$ Mbps and a propagation delay of $50$ ms. Three flows share simultaneously the link: an unresponsive UDP flow with a constant sending rate of $500$ kbps, a flow running the DCCC algorithm and a HSTCP~\cite{hstcp} flow (we provide test cases with TCP New Reno and TCP Westwood in Appendix~\ref{app:sim}). The  delay threshold of the DCCC algorithm has been set to $100$ ms.

\begin{figure}
\centering
\includegraphics[scale=0.18]{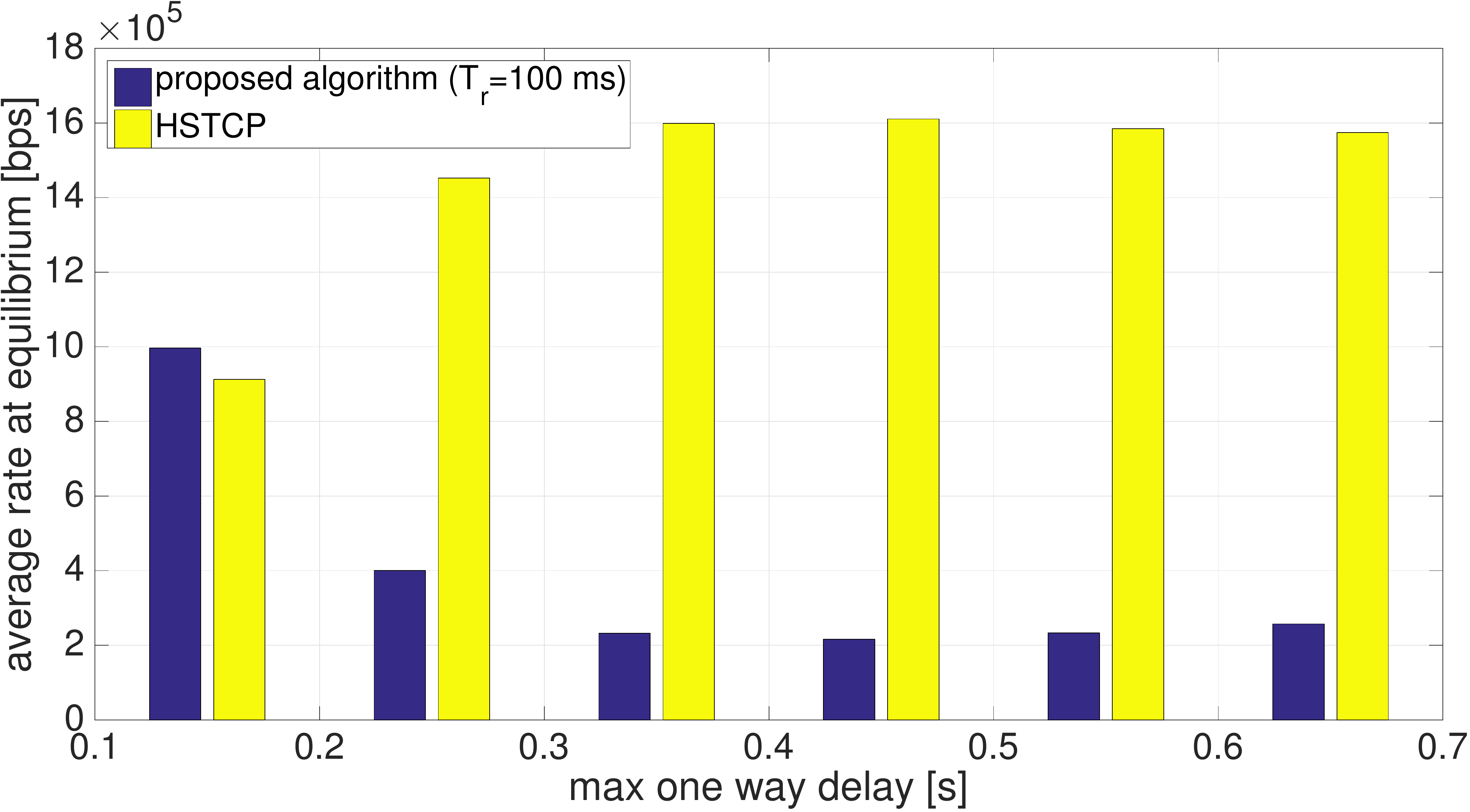}
\caption{Average rate at equilibrium of our algorithm and HSTCP when competing for a bottleneck for different drop tail buffer size.}
\label{TCP}
\end{figure}

We run simulations for different droptail buffer size ranging from $30$ to $180$ packets (corresponding roughly to $100$ ms and $600$ ms of maximum queueing respectively). The simulation results in Fig.~\ref{TCP} show the average rate at equilibrium for the DCCC and TCP algorithms. We can see that the level of fairness against TCP depends on the buffer size. This dependency is caused by the delay-based part of the congestion algorithm, since the buffer size has an impact on the experienced delay and therefore on the rate at the equilibrium.
In the presence of small buffers, our algorithm reaches a higher rate at equilibrium than TCP one. This is due to the fact that the loss-based part of DCCC is more aggressive than the TCP congestion control. On the other hand, in the case of large buffer size, the DCCC algorithm suffers against TCP.  However, due to the non-linearity of the DCCC penalty function, the DCCC flow is protected, and it never starves, reaching the expected lower bound rate of $h/\beta$ ($h/\beta=200$ kbps in the simulations). By modifying the value of $h$, we can tune the guaranteed rate that is reached in large delay environments and therefore we can limit performance degradation when competing with TCP. 
In conclusion, when sharing the bottleneck link with TCP, the DCCC algorithm cannot guarantee TCP fairness, but it still compares favorably with respect to other delay-based algorithms proposed in the literature~\cite{bittorrent,varun,dflow}, which are note able to guarantee a lower bound on the delay-based sending rate.

\subsection{Comparison with Other Congestion Control Algorithms}
We now conduct some experiments to compare our algorithm with other delay-based congestion controllers. We concentrate on the behavior of the algorithms when they operate in delay-based mode and not on how they perform in lossy environments.
We consider two candidate algorithms of the IETF RMCAT
(RTP Media Congestion Avoidance Techniques) Working Group~\cite{rmcat}: NADA and the GCC. NADA is implemented according to the description of the IETF draft~\cite{nadadraft} (version 3) and the scientific publicatio~\cite{nada}; the GCC instead is implemented following the description of the IETF draft~\cite{google-draft} (version 2) and the detailed analysis of the scientific publications~\cite{varun,decicco1,underGCC}. However, both algorithms are moving targets, and the implementation follows the description found in the cited documents. 
To have a more exhaustive evaluation we also include the LEDBAT algorithm in the comparison results. 
For a more detailed discussion of these congestion control algorithms we refer the reader to Section \ref{related}, while for a detailed discussion on the implementation and parameters setting of these algorithms we refer the reader to Appendix~\ref{app:imp}.

\begin{figure}
\centering
\includegraphics[scale=0.17]{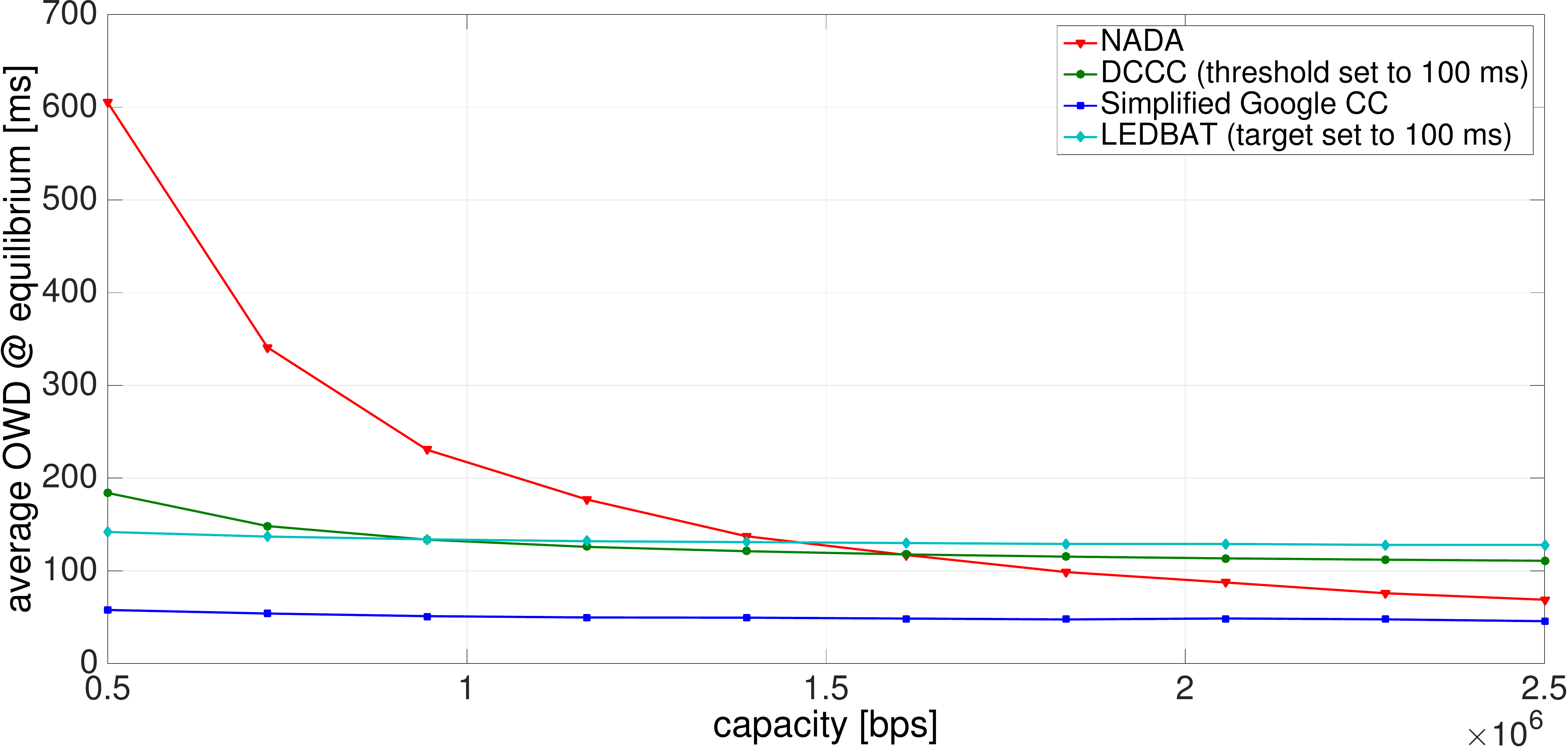}
\caption{Average OWD at equilibrium for the NADA, the simplified version of the GCC and the DCCC algorithms, in a single link topology.}
\label{cmpNADA}
\end{figure}

We consider a single link topology. With a varying channel capacity and propagation delay of $25$ ms. We impose a threshold delay of $100$ ms for the DCCC algorithm and a target parameter of $100$ ms for the LEDBAT protocol. In Fig.~\ref{cmpNADA}, we can see a comparison of the average self-inflicted delay at equilibrium for NADA, GCC, LEDBAT and the DCCC algorithm. 
As ca be observed NADA, LEDBAT and DCCC show a different self-inflicted delay versus bottleneck capacity behavior.
NADA is the one that shows the highest delay variation while LEDBAT is the most stable of the three, whereas DCCC shows an intermediate behavior.
A large self-inflicted delay variation means that the flow is extremely elastic and can achieve a relatively high rate even when a competing flow makes the experienced delay increase to extremely large values. On the other hand, a small variation means that the delay at equilibrium is almost independent of the sending rate, but the flow might starve if another flow in the network forces the queuing delay to increase above this value. The type of behavior that is preferable depends on the type of data that the congestion control has to handle. LEDBAT is preferable for low-priority background traffic, while a behavior similar to NADA is preferable if the goal is to avoid starvation in the presence of an externally imposed high delay.

\begin{figure}
\centering
\includegraphics[scale=0.17]{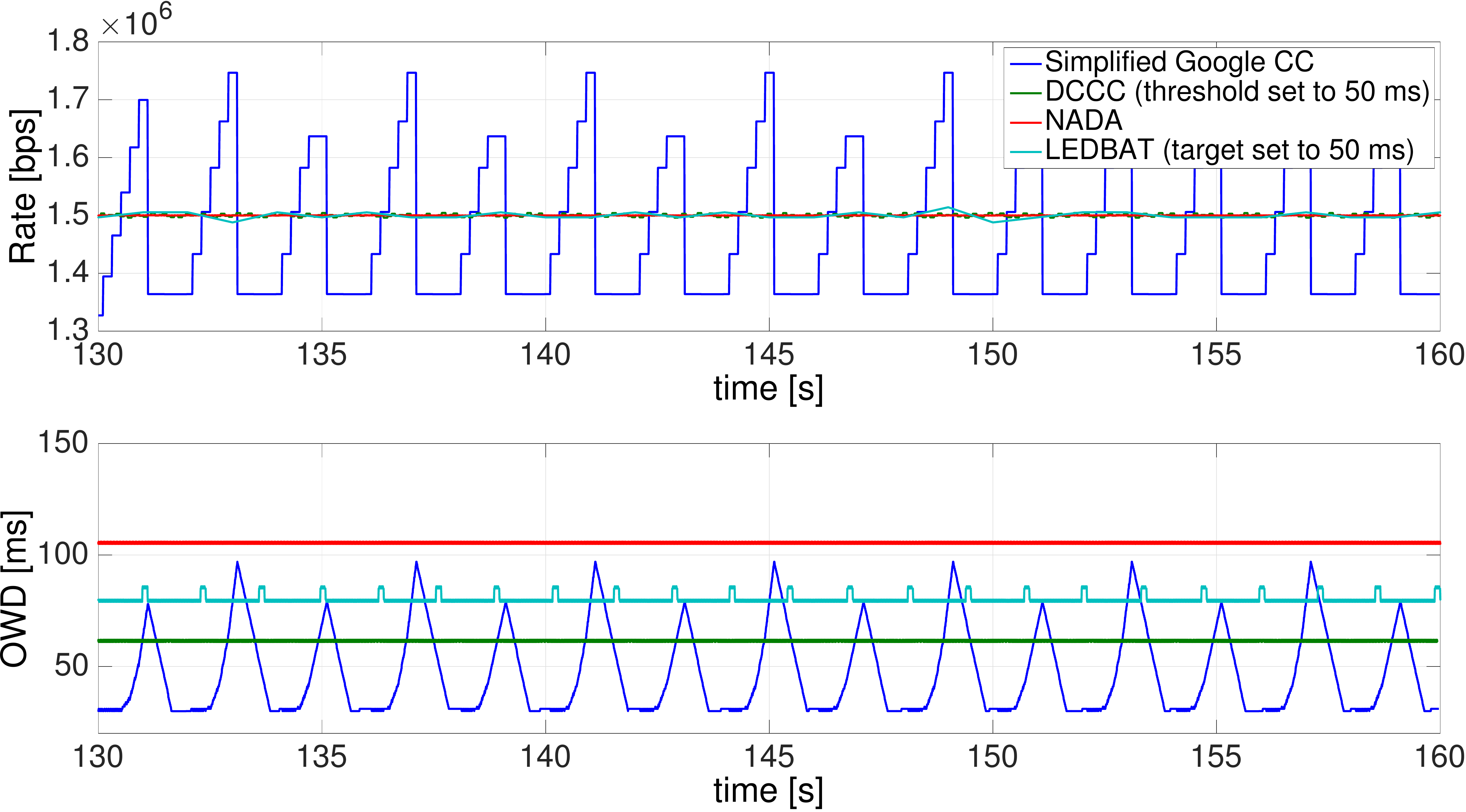}
\caption{Rate and OWD evolution at equilibrium for the NADA, the simplified version of the GCC and the DCCC algorithms, in a single link topology.}
\label{cmpGCC}
\end{figure}

A different discussion has to be made for what concerns the GCC algorithm. 
GCC uses OWD variations and not absolute OWD measurements to adapt the sending rate. Moreover, the GCC keeps increasing the rate until the derivative of the  queueing delay reaches a threshold; at that point, it decreases the rate until the queueing derivative is smaller than a second negative threshold. As a result, it is impossible for GCC to reach an equilibrium rate: it inevitably oscillates between an overestimation and an underestimation of the capacity, therefore the experienced delay also oscillates.
Even if GCC, tested rates, as shown in Fig.~\ref{cmpNADA}, achieves a lower average self-inflicted delay for all the tested rates, the time variability of the OWD of the GCC is relatively large with respect  to the other algorithms in the comparison, thus the maximum experienced delay is much higher than the average one (see Fig.~\ref{cmpGCC}).
In Fig.~\ref{cmpGCC}, we show a comparison of the proposed DCCC, NADA, LEDBAT and our simplified version of GCC when flows stream through a single link of $1.5$ Mbps with a propagation delay of $25$ ms. We focus on the qualitative behavior of the algorithms. DCCC, NADA and LEDBAT use absolute delay measurements and achieve a constant rate and a constant delay. In contrast, the simplified GCC is not able to provide a stable rate at equilibrium since it uses delay variations to compute the sending rate. Using only delay variations to coordinate users, is certainly a good method to mitigate latecomers' advantage and endpoint synchronization requirements. However, this method cannot provide any information of the absolute delay value, which is undoubtedly a key component in users' Quality of Experience. At the current state it is not clear which of the two types of algorithm can lead to better overall performance. 
Finally, we point out that the results shown here are not enough to provide an objective ranking of the algorithms. Rather, they are meant to reveal their main limitations and  advantages.

\section{Related Work}\label{related} 
Among the delay-based congestion control algorithms proposed in the literature, the most similar ones with our controller are the FAST TCP~\cite{FASTTCP},   the CDG~\cite{CDG}, and the LEDBAT~\cite{ledbat}. 

The FAST TCP algorithm is a delay-based  congestion control based on the TCP window-based congestion control. This  allows the FAST TCP protocol to be  easily implemented but also prevents it to be suitable for real-time communications. Moreover, the algorithm adopts the RTT instead of the OWD as delay measure. Therefore, the controller cannot link the delay measure with an overall end-to-end  threshold delay imposed by  the application requirements. 

The CAIA's delay gradient (CDG) congestion control, which is inspired by one of the earliest works on delay-based congestion control~\cite{jaindel},  uses the RTT variations to control the sending rate. By using relative variations of the RTT, the algorithm does not require neither the estimation of the propagation delay nor the use of any fixed threshold. However, only delay variations are considered by the algorithm  and therefore there is no  control on the total delay allowed by   real-time communications.

Finally, the  LEDBAT is a quite recent delay-based congestion control mainly aimed for not prioritized flows. In particular, the controller targets to keep a constant queueing queueing delay at the buffer of the bottleneck link preventing the buffer length to increase indefinitely. In this way, other (possibly more important) flows sharing the same bottleneck cannot go in starvation mode. However,   LEDBAT  suffers from intra-protocol unfairness and it starves when competing against regular TCP flows (see~\cite{outta,bittorrent} for a detailed evaluation), therefore it cannot be considered as viable controller for low-delay (but still important) communications. 

In~\cite{CxTCP}, the authors present an interesting approach to solve the complex problem of loss-based and delay-based coexistence.  The work proposes a modification of the TCP algorithm, called Cx-TCP, with a backoff probability that is a linearly increasing function of the experienced queueing delay for small delay values, and it is null for  large   delay values. This allows the system to do not back off  when  delays are due to the coexistence with loss based flows (i.e., when large delay are experienced), avoiding starvation states. The resulting controller achieved good performance but, similarly to FAST TCP, being  a modification of the TCP protocol, its properties (e.g., packet retransmissions) are not ideal for real time applications such as video conference applications.

More recent solutions have been proposed in the framework of RMCAT: a working group of the IETF~\cite{rmcat} which aims to standardize a congestion control algorithm for real-time media applications  that is able to work in different network environments.  Having  the same final target of our proposed controller, the algorithms proposed in this group share similar objectives with our controller. One proposal is the NADA controller~\cite{nadadraft}, which uses the estimate of the  actual propagation delay of the route as delay penalty and packet losses and an Early Congestion Notification (ECN)~\cite{ecn} marking for the loss penalty. The algorithm is able to maintain a consistent sending rate when competing against TCP flows and it    maintains low end-to-end delays for medium- high- bottleneck link capacities. However, the controller suffers from large self-inflicted delays for low bottleneck link capacities.  It is worth noting however that NADA is still ongoing work and according to the last version of the IETF draft the update rate equation has remarkably changed from the work~\cite{nada}. In particular, the self-inflicted delay of the control algorithm in low capacity environments has been significantly reduced with the new design.

Another proposal in the RMCAT working group is the GCC algorithm~\cite{google-draft}. This controller is mainly composed of two subparts:  the loss-based part is strongly based on the TFRC~\cite{tfrc} throughput equation, while the  second subpart consists of a delay-based congestion control algorithm. The delay-based algorithm in this case is substantially different from the other proposals. The  GCC uses the queuing delay derivative, rather than its absolute value, to tune the sending rate. Since it is not required to estimate the absolute propagation delay the advantage of this method is that the endpoints do not have to be synchronized. The main issues of this delay-based algorithm is however the poor intra-protocol fairness~\cite{decicco1,varun}. For what concerns the loss-based subpart, it has been shown in~\cite{decicco1} that the original version of the GCC suffers from starvation when competing against TCP flows, However in~\cite{underGCC}, the authors have proposed a modification of the delay-based algorithm that permits to alleviate this problem. 

Finally, DFLOW~\cite{dflow} is another loss- and delay-based congestion control proposed in the RMCAT group. The loss-based subpart of the algorithm basically corresponds to the TFRC rate equation, while the delay-based term   shares some similarities with the LEDBAT algorithm. According to the authors, DFLOW provides only limited throughput when competing against loss-based flows, such as TCP. Another more recent proposal in the RMCAT group is SCREAM which stands for self-clocked rate adaptation for multimedia~\cite{scream}. This solution uses a hybrid loss- and delay-based congestion control aimed at working in wireless LTE scenarios. The algorithm conforms to the packet conservation principle since it sets an upper limit on the amount of data that is allowed in the network, similarly to how window based algorithms such as TCP work. However, according to some publicly available results~\cite{slidesSCREAM}, SCREAM achieves a low throughput when competing against TCP flows in the scenario of long buffers.

In summary, several hybrid loss- and delay-based congestion control algorithms have been proposed for real-time communication. While all nicely performing in terms of experienced delay, loss-based coexistence and delay estimation are the two main  open challenges.   Our proposed  controller  alleviates both of them achieving $i)$  improved performance when sharing the network resources with loss-based flows; $ii)$  preserving intra-protocol fairness in critical scenarios.

\section{Conclusion}\label{conc}
In this work,  we have developed and analyzed a novel hybrid delay- and loss-based congestion control algorithm, namely the DCCC algorithm. The proposed algorithm is able to $i)$ maintain  a bounded delay communication if the network conditions allows it; $ii)$ prevent starvation when competing against loss-based flows. Introducing a price  measure based on the interarrival time of the packets, we are able to provide a controller that automatically behaves as  delay-based and loss-based protocol, based on the actual event that triggers the congestion. Moreover, because of the non-linear mapping between the experienced delay and the delay-based congestion signal, the   DCCC algorithm avoids starvation when competing against loss-based flows. The non-linearity mapping also mitigates unfairness problems when the data paths of the users have different propagation delays. Finally, by using the total experienced delay instead of the actual queueing delay, we avoid estimation problems and unfairness issues due to latecomer flows.  The ability to achieve a bounded delay at the equilibrium and the flexibility of being able to not starve against loss-based flows makes the DCCC algorithm a suitable congestion control algorithm to be used for delay sensitive network applications, e.g. video conferencing. As further work, we propose to improve TCP fairness of the DCCC algorithm and the speed of convergence to the equilibrium point.

\appendices

\section{Non-linear Stability Proof}\label{app:proof}
We start by analyzing the users subpart of the system. The time derivative of the storage function in Eq. \eqref{storage1} is given in Eq. \eqref{timeder}.
\begin{figure*}[t]
\normalsize
\begin{IEEEeqnarray}{rCl}
	\dot{V}_{user} & = \ & \frac{1}{\alpha} \delta x^T\left( U'(x) - v(e) - \dot{e} \right)_{x}^+   \IEEEyessubnumber \label{timeder}\\
                		      & \leq \ & \frac{1}{\alpha} \delta x^T\left( U'(x) - v(\hat{e}) \right) -\frac{1}{\alpha} \delta x^T\dot{e} - \frac{1}{\alpha} \delta x^T\left( v(e) - v(\hat{e}) \right)  \IEEEyessubnumber \label{firstdis} \\
               		      & = \ & \frac{1}{\alpha} \delta x^T\left( U'(x) - U'(\hat{x}) \right) -\frac{1}{\alpha} \delta x^T\dot{e} - \frac{1}{\alpha} \delta x^T\left( v(e) - v(\hat{e}) \right)  \IEEEyessubnumber\\
                		      & < \ & \frac{1}{\alpha} \delta x^T\left( U'(x) - U'(\hat{x}) \right) -\frac{1}{\alpha} \delta x^T\dot{e} -  \delta x^T\delta e \IEEEyessubnumber  \label{secdis} \\
                		      & = \ &  \underbrace{\frac{1}{\alpha} \delta x^T\left( U'(x) - U'(\hat{x}) \right)}_\text{$-W_{user}(x)$} + (-\frac{1}{\alpha} \delta y^T\dot{d})   + (-\delta y^T)\delta d \IEEEyessubnumber \label{fin} 
\end{IEEEeqnarray}
\hrulefill
\end{figure*}
It leads to $\dot{V}_{user}<-W_{user}(x) -1/\alpha \delta y^T\delta \dot{d} -\delta y^T\delta d$ which has the form of a passive function similar to Eq. \eqref{eq:pass}.

The result in Eq. \eqref{fin} comes from a first inequality in Eq. \eqref{firstdis}, which holds due to the projection onto the positive orthant of Eq. \eqref{dyna}, and from the second inequality in Eq. \eqref{secdis}, which holds if:
\begin{equation}
\frac{v(e) - v(\hat{e}) }{\alpha}  <\delta e.
\label{price_f}
\end{equation}
This last inequality in Eq. \eqref{price_f} is true if the price function is chosen to be non decreasing with a maximum derivative equal to $\alpha$, which is verified for our price function in Eq. \eqref{del_pen}. It remains to show that $W_{user}(x)$ is a positive definite function, which is needed to use the passivity theorem and eventually prove the stability of the system. The first term in $W_{user}(x)$ is: $- \delta x^T\left( U'(x) - U'(\hat{x}) \right) $, which is always positive due to the concavity of the utility function. 

For the second subpart of the system, namely the dynamics of the links, we now prove that it is also a passive system. We can easily calculate the time derivative of the storage function defined in Eq. \eqref{storage2}:
\begin{IEEEeqnarray}{rCl}
\dot{V}_{link} & = &\delta q^T(y-c)^+_q + \frac{1}{\alpha} (c - \hat{y})^T \dot{q} \IEEEyessubnumber \label{eq:links_1}\\
		      & \leq &\delta q^T(y-c+\hat{y}-\hat{y})+ \frac{1}{\alpha} (c - \hat{y})^T \dot{q} \IEEEyessubnumber \label{eq:links_2}\\
		      & = &\delta q^T(\hat{y}-c) + \delta q^T\delta y + \frac{1}{\alpha} (c - \hat{y})^T \dot{q} \IEEEyessubnumber \label{eq:links_3}\\
		      & \leq &\underbrace{\delta q^T(\hat{y}-c)}_\text{$-W_{link}(q)$} + \delta d^T\delta y + \frac{1}{\alpha} \delta y^T \dot{d} \IEEEyessubnumber \label{eq:links_4}.
\end{IEEEeqnarray}
Where we used the fact that $\delta q=\delta d$ and $\dot{q}=\dot{d}$. Inequality \eqref{eq:links_2} is true due to the projection onto the positive orthant, while inequality \eqref{eq:links_4} holds if $\delta y^T \dot{q} \geq (c - \hat{y})^T \dot{q}$, this can easily be proved in the following way:
\begin{IEEEeqnarray}{rCl}
\delta y^T \dot{q} - (c - \hat{y})^T \dot{q} & = & (y-\hat{y})^T {\delta \dot{q}} - (c - \hat{y})^T \delta \dot{q}\IEEEyessubnumber\\
                             			             & = & (y-c)^T \delta \dot{q}\IEEEyessubnumber\\
                            			             & = & (y-c)^T(y-c)_{q}^+>0.\IEEEyessubnumber
\end{IEEEeqnarray}
The two right most terms in Eq. \eqref{eq:links_4} simplify with the two terms in \eqref{fin} when the two storage function are summed together. The missing part is to show that $W_{link}$ is positive semidefinite, which means $(q-\hat{q})^T(\hat{y}-c) \leq 0$. In order to prove this consider the following deductions. If the link at the equilibrium is fully utilized, the value of the difference  $\hat{y}_l-c_l$ is zero; otherwise it must be negative. At equilibrium a link rate cannot be greater than the link capacity, otherwise it would mean that the queue is growing contradicting the hypothesis of equilibrium. Alternatively, if the link is partially utilized the difference $q_l-\hat{q}_l$ is necessarily non negative, since $\hat{q}_l=0$; otherwise it can be either positive or negative. Combining these two deductions, we observe that the product of $\delta q^T(\hat{y}-c)$ is always non positive. Finally, the global stability of the undelayed control system for the case of no losses is guaranteed, as the system is the combination of two passive systems \cite{passbook}, and the sum of the two storage functions is a Lyapunov function for the entire dynamic system.

\section{Parameters Setting}\label{app:param}
\textcolor{black}{
In this section, we provide guidelines for setting the parameters used in the DCCC algorithm. For the sake of clarity, in Table \ref{tab:sens} we further list the effects of each parameter on the system performance.}

\begin{table}[]
\centering
\textcolor{black}{
\caption{Sensitivity to parameters}
\label{tab:sens}
\begin{tabular}{c|cccc}
Parameter $\rightarrow$\\-----------------------\\Performance\\affected $\downarrow$                                                                                                                   & $h$ & $T_r$  & $\beta$ & $k_r$         \\ \hline \hline
stability sensitivity                                                                                                    & low$(\downarrow)$    & extremely low$(\downarrow)$    & high$(\downarrow)$    & high$(\downarrow)$          \\ \hline
\begin{tabular}[c]{@{}c@{}} convergence speed\end{tabular}                              & medium$(\uparrow)$   & medium-high$(\downarrow)$   & low$(\uparrow)$   & high$(\uparrow)$          \\ \hline
\begin{tabular}[c]{@{}c@{}} threshold\\overshoot\\at equilibrium \end{tabular}      & high$(\uparrow)$   & medium$(\uparrow)$ & high$(\downarrow)$    & null          \\ \hline
\begin{tabular}[c]{@{}c@{}} value of delay at\\equilibrium \end{tabular}              & medium$(\uparrow)$ & high$(\uparrow)$   & medium$(\downarrow)$  & null          \\ \hline
\begin{tabular}[c]{@{}c@{}} minimum\\delay-based\\sending rate \end{tabular}  & high$(\uparrow)$   & null   & high$(\downarrow)$    & null          \\ \hline
\begin{tabular}[c]{@{}c@{}} loss-based\\equilibrium\\sending rate\\(small buffers) \end{tabular}  & high$(\uparrow)$   & null     & null   & - \\ \hline \hline
\end{tabular}}
\end{table}

The parameters that strongly affect the stability of the algorithm and the time to convergence are $\beta$, $k_r$ and $T_s$. To guarantee the stability, $\beta$ and $k_r$ are fixed according to the conducted stability analysis (Subsection \ref{secdelay}).
Since the value of $M$ cannot be known in a realistic scenario we consider its value to be equal to $1$.
In our implementation, we set first $T_s=RTT$, then we set the gain $k_r$ in order to have a cross frequency smaller than $1/T_s$ and avoid aliasing effects. Finally, we tune $\beta$ such that the zero and the pole of Eq. \eqref{TF_2} are located at frequencies smaller than the cross frequency.
This guarantees that the contribution of the zero-pole pair (in terms of magnitude gain and phase delay) cancels out at the cross frequency, simplifying the allocation of this last quantity. Finally, we set the cross frequency to a value that is sufficiently low to ensure the stability of the system, but still large enough to reduce the time to convergence. Having a value of $k_r=1/(2.5RTT)$ and a value of $\beta$ equal to $0.1$ provides a good tradeoff between stability and speed. Note that the value of $T_r$ affects the speed of convergence because $T_r$ affects the RTT  which ultimately affects the gain $k_r$. Having a low threshold delay tends to make the controller faster.

The value of $h$ affects the aggressiveness of the controller since the value of the price at the equilibrium, e.g., the overshoot of the delay threshold in a delay-based scenario, is proportional to $h$.
The value of $h$ affects also the minimum rate that is achieved in the case of extremely large delay penalties. This minimum delay-based rate is equal to $h/\beta$, and ideally, a larger rate is better.
The tuning of the parameter  $h$ represents the following tradeoff: increasing $h$ leads to a larger guaranteed rate for the delay-based part, but also to larger delays at equilibrium when operating in delay mode and outclassing the TCP in the presence of short buffers.
Finally the parameter $h$ can also be used to assign different levels of importance to the different flows, in this case we denote with $h_r$ the value of the parameter for user $r$. DCCC flows with the same delay threshold and the same propagation delay, will achieve different rates at equilibrium depending on the value of $h_r$: the higher the value of $h_r$ the higher the sending rate. In our implementation we set $h_r=h\ \forall r$ since we are not interested in prioritizing flows. We set $h=20$ kbps since it offers a fair balance of the aforementioned tradeoff.

The value of $T_r$ mainly affects the delay at equilibrium (the threshold delay affects also the delay overshoot and the convergence speed, however these are rather side effects that appear because  $T_r$ affects the RTT at equilibrium). As a result, the main aspect that has to be taken into account when setting $T_r$ is the OWD at equilibrium.
In general, by properly tuning $T_r$, the application can achieve larger rates, at the price of higher delays. This is actually the tradeoff faced by realtime applications: higher delay and likely higher rate or lower delay and likely lower rate. Since $T_r$ does not affect strongly the stability, it can be set freely  to achieve the right tradeoff between achieved rate and experienced delay.
It is worth noting that $i)$ to measure the right OWD, two endpoints need to be synchronized; $ii)$  to fully utilize the channel,  the minimum OWD needs to be larger then $T_r$.  We can then list the different scenarios to provide a complete solution to choose the threshold delay:
\begin{enumerate}
\item Nodes are synchronized, the minimum OWD is approximately known (or upper bounded): Set $T_r$ to a value that provides a good quality of service to the final user. If this value is smaller than the estimated minimum OWD, $T_r$ can be set equal to the approximate minimum OWD. Note that in the latter case  there is no way to provide a good quality of service to the user.
\item Nodes are synchronized, the minimum OWD is not known: Either set $T_r$ to a value that provides a good quality of service to the final user, and possibly does not lead to utilizing the channel efficiently  or set $T_r$ to the minimum observed OWD, assuring that the channel is used completely but possibly at the price of latecomers' advantage issues, that are common to many other delay-based algorithms.
\item Nodes are not synchronized. In this case, since there is no way to have a meaningful measure of the OWD, the only feasible choice is to set $T_r$ to the minimum OWD observed.
\end{enumerate}
In practice, the Network Time Protocol (NTP) can be used to synchronize the endpoints. If nodes are not synchronized, the effect of desynchronization affects the algorithm performance in a gradual manner. More precisely, when the endpoints are desynchronized by an amount of time $\Delta T_{\text{sync}}$, the effect on the congestion control algorithm is the same of adding the value $\Delta T_{\text{sync}}$ to the delay threshold $T_r$. Thus, the effect of the desynchronization stays negligible as long as $\Delta T_{\text{sync}}\ll T_r$.

In this section we have analyzed in depth the different effects that the parameters have on the algorithm performance, providing general guidelines on how they should be set.

\section{Practical Implementation}\label{app:imp}
In this section we first describe more in detail the implementation of the DCCC algorithm, as well as the one of the algorithms used for the comparison.

\subsection{DCCC Implementation}
We implement our congestion control algorithm on top of the UDP protocol, the size of the packets is set to $1$ KB of data plus the size required by the protocol headers, DCCC (40 B) and UDP-IP (30 B), thus the total size is 1094 B. 
We provide here a brief list of steps which describe how the controller can be implemented:
\begin{description}
\item[\emph{On media packet sending}] \hfill \\ Add to the packet header: the current timestamp, the current sending rate and the current RTT.
\item[\emph{On media packet received}] \hfill \\ Get and store: the current OWD, the interarrival time with respect to the previous packet, the sending rate and the RTT reported in the media header.
\item[\emph{On feedback packet sending}] \hfill \\ Average the OWD, the interarrival time of the packets and the sending rate, for the packets received since the last feedback sending event. Send this information to the sender adding the current timestamp. Schedule the next feedback sending in one RTT.
\item[\emph{On feedback packet received}] \hfill \\ Get the information reported in the feedback packet: the current OWD ($e_r$), the received sending rate ($x_r(t-RTT_r)$), the average interarrival time ($\Delta t^a$). Then compute the RTT based on the OWD and the backwards OWD from the feedback packet. compute Eq. \eqref{finalform} and update the sending rate.
\end{description}
It is worth noting that there is no unique way to implement the algorithm. The computation of the received rate can be done entirely at the receiver side, or alternatively all the operations can be left at the sender side by using a per packet feedback.
Finally, in our implementation the timestamps have an accuracy of $1$ ms. Since time accuracy is limited in real systems, it might be challenging to achieve a sufficient precision for the interarrival time measurement at high sending rates. Though we do not tackle this specific scenario but rather typical real-time applications that do not send data at extremely high speeds, executing the rate update on a fixed $T_S$ basis and not on a per packet basis improves the resilience of our algorithm to timestamp accuracy issues. In fact at high rates we can average the interarrival time over a large number of packets, increasing ultimately the measurement accuracy. In our implementation for example, the interarrival time of the packets is averaged in one RTT window ($T_S=\text{RTT}$), thus we simply need to count the number of packets received in one RTT window in order to estimate the interarrival time.

\subsection{Comparison Algorithms Implementation}
\textcolor{black}{
We briefly describe here the NADA, GCC and LEDBAT algorithms implementations.}

\textcolor{black}{
The NADA algorithms implementation follows the description of the IETF draft~\cite{nadadraft} (version 3) and of the conference paper \cite{nada}, with the exception of the shaping buffer part, which is not needed since we assume an ideal data generator. All the parameters are set accordingly to the description present in the simulation section of the conference paper. In particular the value of the reference delay $x_{ref}$ is set equal to the minimum OWD observed during one simulation episode. We then set $R_{MAX}=6$ Mbps and  $R_{min}=0.1$ Mbps; the smoothing parameter corresponds to $\alpha=0.001$; and the prediction parameter is set $\tau_0=0.1$.}

\textcolor{black}{
In the GCC IETF draft, the algorithm description is bounded to a realistic video encoding system, while our simulations assume an ideal data generator.
We thus built a simplified Google congestion controller that uses an ideal data generator, following the description of the IETF draft~\cite{google-draft} (version 2) and the publications \cite{decicco1,varun}. This permits to avoid issues such as data bursts or rate mismatch, possibly caused by the use of real live encoders, which may lead to an unfair comparison among controllers. The mathematical model of our simplified version of the Google algorithm is however the same of the original controller version, with some minor modifications involving the delay filtering. The original draft uses a Kalman filter to estimate the queuing delay variation from a set of packets that are sent simultaneously. In our implementation, the packets are not sent in bursts and the delay measurement made on a single packet is sufficiently accurate to be used as a reliable measurement. Finally, we kept the threshold parameter $\gamma$ fixed. The adaptation of this parameter seems to be extremely important to improve TCP coexistence \cite{underGCC}. However, in our comparison we focus only on the self-inflected delay with no loss-based cross traffic, so that a fixed parameter does not compromise the comparison (the parameter adaptation has been introduced in the version 2 of the draft as suggested in~\cite{underGCC}).
The parameters are set according to the Google draft \cite{google-draft} (version 2) and the descriptions found in \cite{varun,decicco1,underGCC}. We set then $\eta=1.2$, $\alpha=0.9$, $\gamma=0.4$ ms and the REMB sending interval is set to 1 s.}

The LEDBAT is implemented as described in the IETF document \cite{ledbat}. In particular the value of the Maximum Segment Size (MSS) is equal to the size of the packets, which is $1094$ B, the gain parameter is set to $0.5$, the allowed increase is set to one MSS, the minimum congestion window to $2$ MSS. Finally, the target parameter is set within the range $50$-$100$ ms (actual value specified in the different simulations) ms consistently with the guidelines of the LEDBAT and aligned with the settings of the DCCC threshold.

Finally, note that the total size of the packets for the three algorithms is equal to $1094$ B, while the size of the packets for the UDP streams used in the different simulations is equal to $1054$ B.

\section{Supplementary Simulation Results}\label{app:sim}

\subsection{Bandwidth Variations}
\begin{figure}
\centering
\includegraphics[scale=0.10]{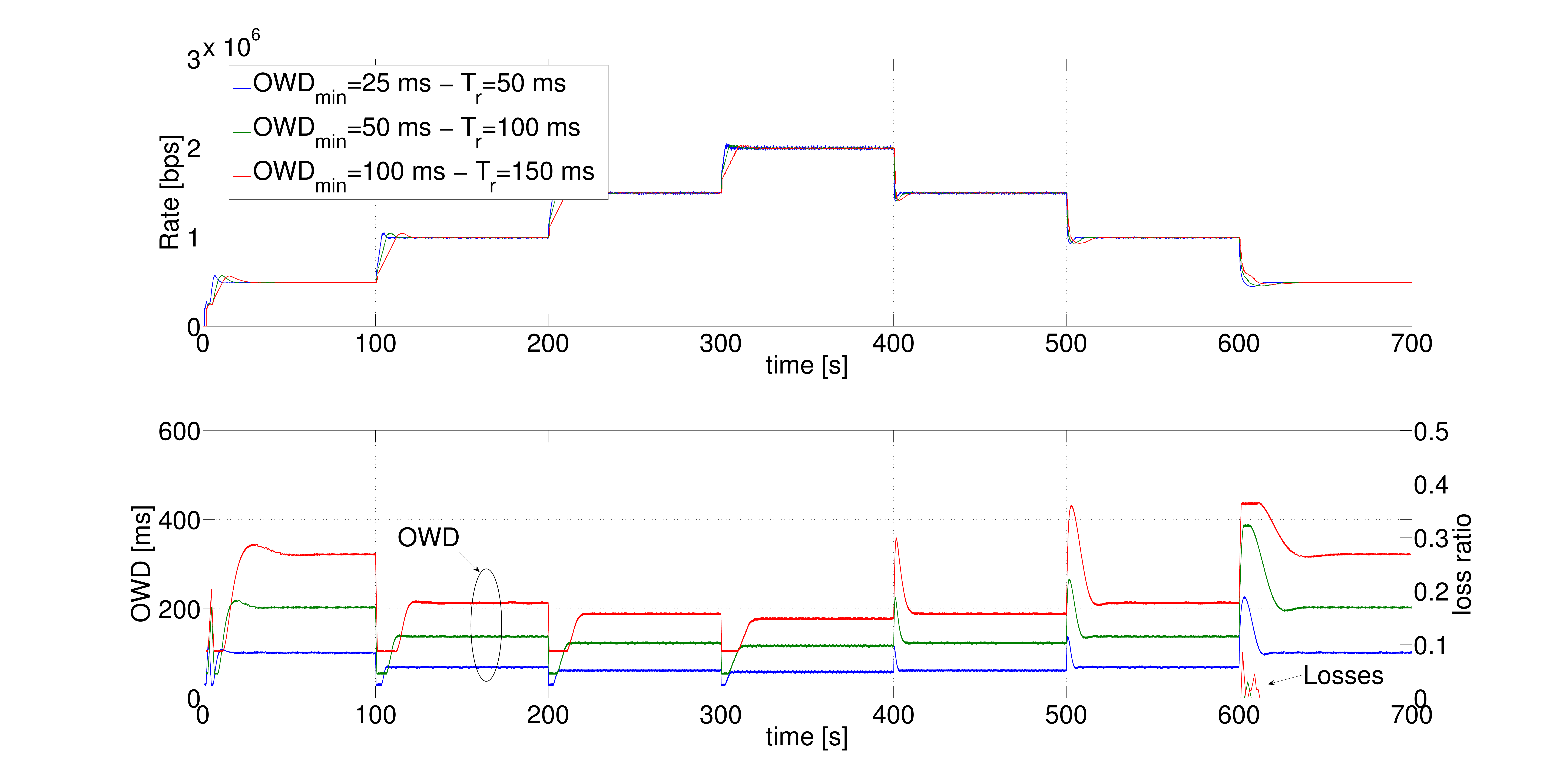}
\caption{Evolution of the sending rate over a stair-case available bandwidth.}
\label{stair2}
\end{figure}

As first scenario, we simulate a single link topology with an available bandwidth varying over time.
To simulate this dynamic scenario, we consider an unresponsive (i.e., constant bitrate) UDP flow sharing the bottleneck link with the DCCC flow. While the total capacity of the bottleneck link is fixed to $2.5$ Mbps, the rate of the UDP flow varies between $500$ kbps to $2$ Mbps. Buffers implement a droptail policy with a maximum buffer length of $100$ packets, which corresponds to a maximum queueing delay of $330$ ms. We run three different simulations with different propagation delays of the bottleneck link, namely $25$, $50$ and $100$ ms, and different delay thresholds, namely $50$, $100$ and $150$ ms.
In Fig. \ref{stair2} we provide the results for this scenario. Simulation results show that the DCCC algorithm always fully utilizes the available bandwidth left by the UDP flow.  As expected the algorithm converges at a delay that overshoots the imposed delay threshold of an offset inversely proportional to the equilibrium rate: the lower the rate the larger the expected delay at equilibrium.

\subsection{Noisy Delay Measurement}
\begin{figure}
\centering
\includegraphics[scale=0.1]{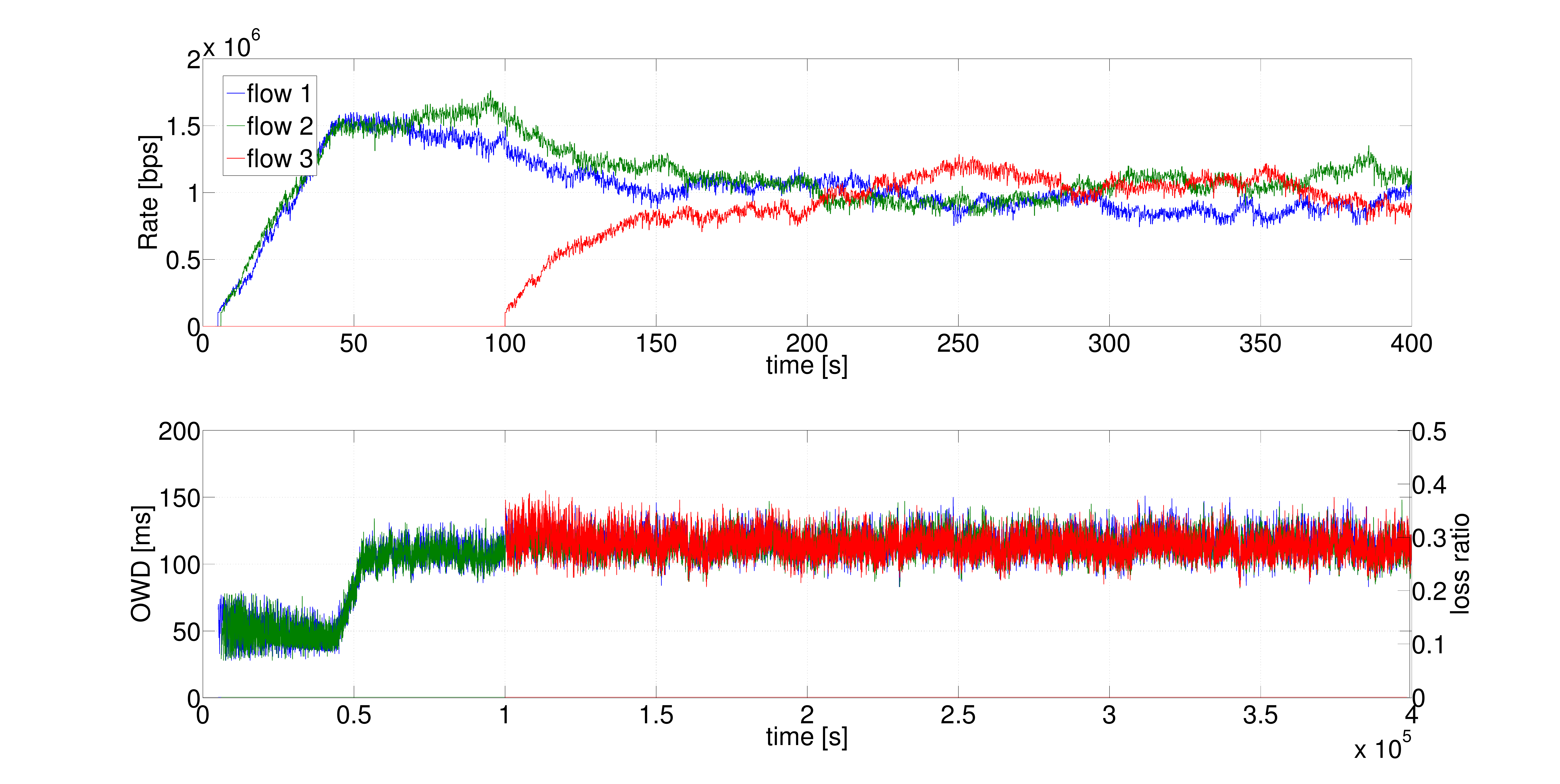}
\caption{Sending rate and OWD for the three DCCC flows sharing a common bottleneck link when the equilibrium is driven by the experienced delay, with noisy delay measurements.}
\label{jitter}
\end{figure}

\begin{figure}
\centering
\includegraphics[scale=0.1]{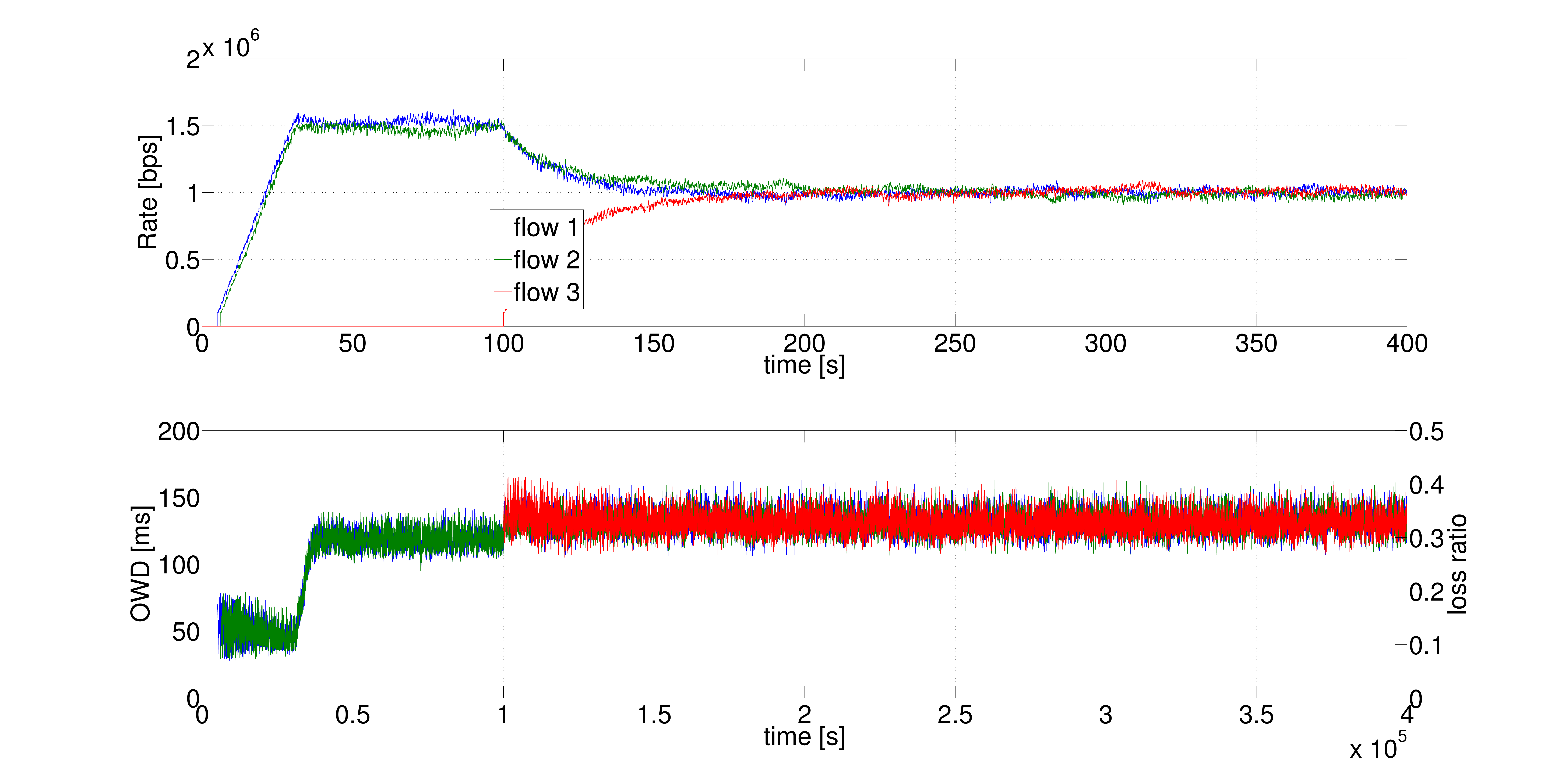}
\caption{Sending rate and OWD for the three DCCC flows sharing a common bottleneck link when the equilibrium is driven by the experienced delay, with noisy delay measurements and smoothed receiving rate estimation.}
\label{smooth}
\end{figure}

We now analyze the performance of the proposed algorithm in the case of noisy delay measurements. In this scenario, three flows share a common bottleneck of $3$ Mbps with a maximum queuing delay of $300$ ms and a minimum propagation delay of $25$ ms. Flows $1$ and $2$ start at the beginning of the simulation, whereas flow $3$ starts after $260$ s. All the flows have a threshold delay of $100$ ms. In order to introduce the noise to the delay measurements, every packet is delayed by a uniformly distributed random period in the range of $[0,50]$ ms.\footnote{The random delay is increased in case it would lead to packet disordering.}
Results are shown in Fig. \ref{jitter}. We can see that the proposed algorithm is able to correctly operate in noisy delay scenarios even without modifications. To reduce the noise that affects the sending rates, we further apply a smoothing operation to the receiving rate estimation as follows
\begin{equation}
\tilde{x}_r^{recv}=\alpha \tilde{x}_r^{recv} + (1-\alpha) x_r^{recv}.
\label{eq:smooth1}
\end{equation}
Fig. \ref{smooth} provides the results of the proposed algorithm with the addition of the smoothing operation of Eq. \eqref{eq:smooth1} with $\alpha=0.5$. Simulations results show a good improvement in the rate stability and prove the robustness of the proposed algorithm to noisy delay measurements.

\subsection{TCP Coexistence}
\begin{figure}
\centering
\includegraphics[scale=0.18]{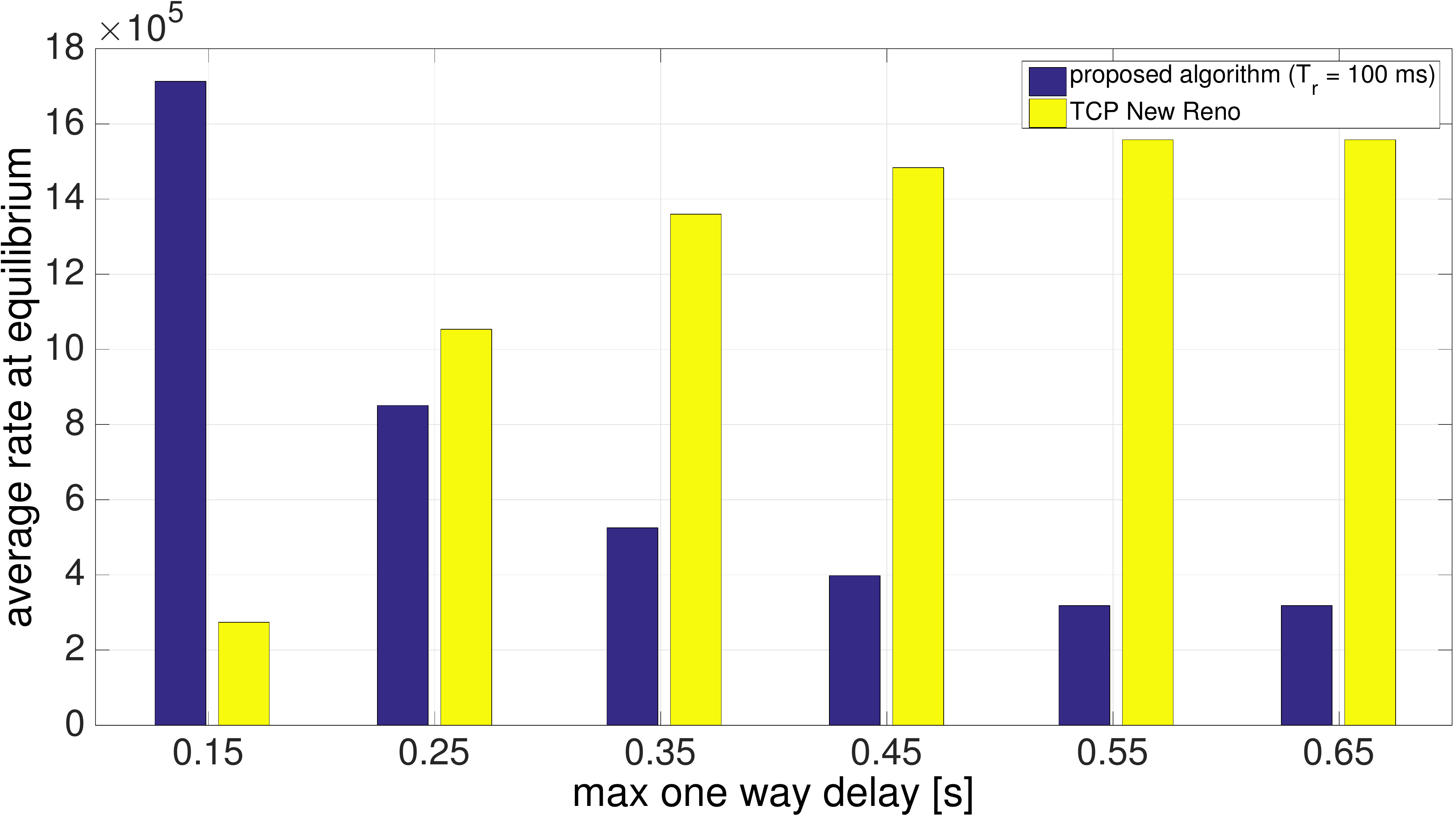}
\caption{Average rate at equilibrium of our algorithm and TCP New Reno when competing for a bottleneck for different drop tail buffer size.}
\label{newreno}
\end{figure}

\begin{figure}
\centering
\includegraphics[scale=0.18]{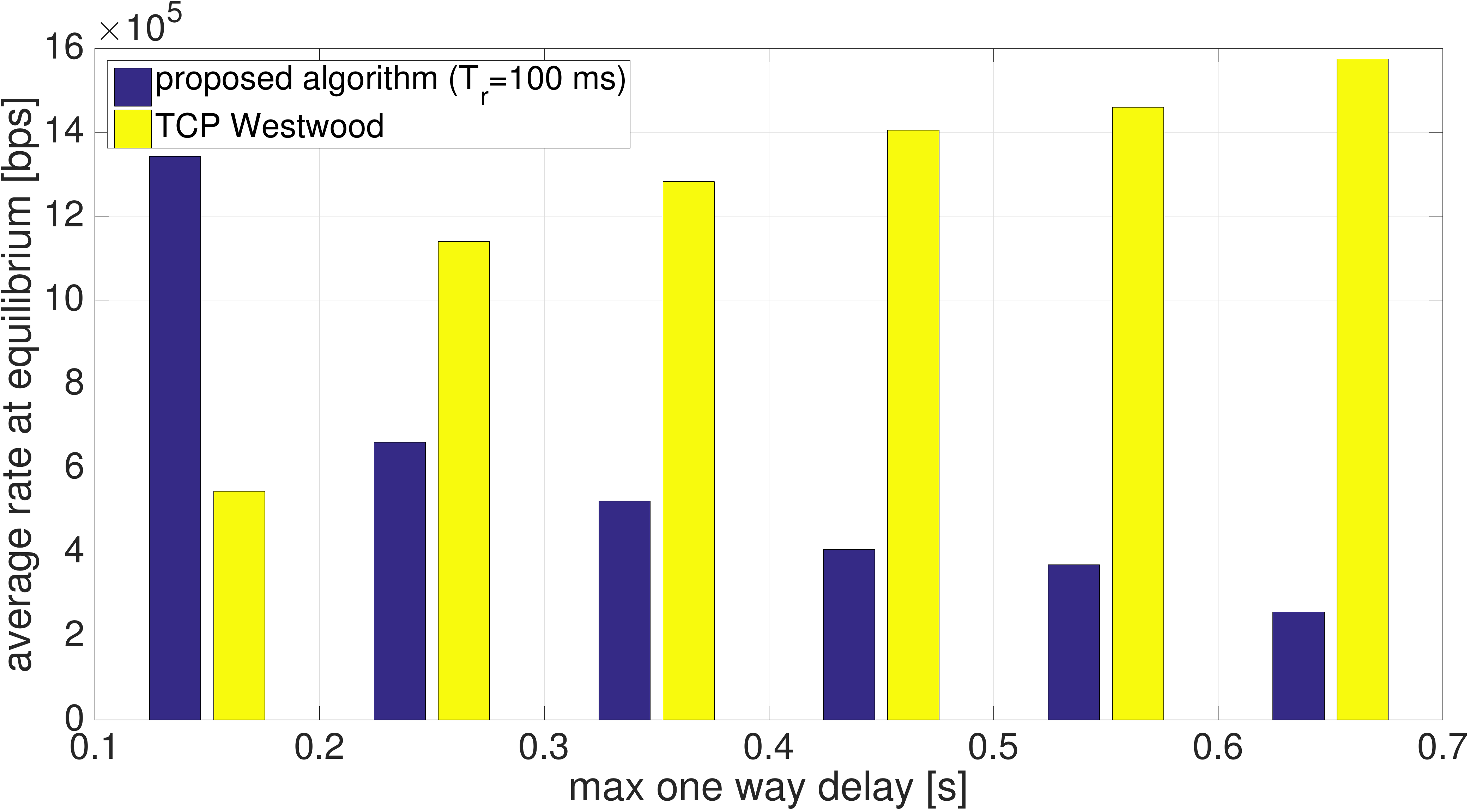}
\caption{Average rate at equilibrium of our algorithm and TCP Westwood when competing for a bottleneck for different drop tail buffer size.}
\label{westwood}
\end{figure}

We conduct here the same type of experiments of Subsection \ref{sec:tcp} but with different TCP flavors, namely TCP New Reno and TCP Westwood. The settings of the simulations are exactly the same of those described in Subsection \ref{sec:tcp}. In Fig. \ref{newreno} and \ref{westwood} are shown the results for the TCP New Reno and TCP Westwood respectively. As can be seen, though the aggressiveness of the TCP flows slightly changes, in all cases our algorithm avoids starvation by sending data at a rate equal to $h/\beta$, which in this case corresponds to $200$ kbps, when buffers are excessively long.


\section*{Acknowledgment}
This work has been partly supported by the Swiss Commission for Technology and Innovation under grant CTI-13175.1 PFES-ES and by the Swiss National Science Foundation under grant CHISTERA FNS 20CH21\_151569.





%

\bibliographystyle{./IEEEtran}
\bibliography{./IEEEabrv,./draft_jrnl}

%
%

%








\end{document}